\documentclass[aps,twocolumn,superscriptaddress,groupedaddress]{revtex4-1}
\usepackage{amsmath}
\usepackage{subfigure}
\usepackage{graphicx}
\usepackage{dcolumn}
\usepackage[colorlinks=true, citecolor=blue,linkcolor=blue,urlcolor=blue
]{hyperref}
\usepackage{color}
\usepackage{bm}

\newcommand{\beq}{\begin{equation}}
\newcommand{\eeq}{\end{equation}}

\begin{document}
\author{Muhammad Asjad}
\affiliation{School of Science and Technology, Physics Division, University of Camerino, Camerino (MC), Italy}

\author{David Vitali}
\affiliation{School of Science and Technology, Physics Division, University of Camerino, Camerino (MC), and INFN, Sezione di Perugia, Italy}
\title{Reservoir engineering of a mechanical resonator: generating a macroscopic superposition state and monitoring its decoherence}

\begin{abstract}
A deterministic scheme for generating a macroscopic superposition state of a nanomechanical resonator is proposed. The nonclassical state is generated through a suitably engineered dissipative dynamics exploiting the optomechanical quadratic interaction with a bichromatically driven optical cavity mode. The resulting driven dissipative dynamics can be employed for monitoring and testing the decoherence processes affecting the nanomechanical resonator under controlled conditions.
\end{abstract}

\maketitle

\section{Introduction}

Quantum reservoir engineering generally labels a strategy which exploits the non-unitary evolution of a system in order to generate robust quantum coherent states and dynamics \cite{Diehl2008}. The idea is in some respect challenging the intuitive expectation that in order to obtain quantum coherent dynamics one should guarantee that the evolution is unitary at all stages. Due to the noisy and irreversible nature of the processes which generate the target dynamics, strategies based on quantum reservoir engineering are in general more robust against variations of the parameters than protocols solely based on unitary evolution~\cite{Diehl2008,Verstraete2009}. A prominent example of quantum reservoir engineering is laser cooling, achieving preparation of atoms and molecules at ultralow temperatures by means of an optical excitation followed by radiative decay~\cite{Wineland1978}. The idea of quantum reservoir engineering has been formulated in Ref.~\cite{Poyatos1996}, and further pursued in Ref.~\cite{Carvalho2001}. Proposals for quantum reservoir engineering of many-body systems have been then discussed in the literature~\cite{Diehl2008,Verstraete2009} and first experimental realizations have been reported~\cite{Syassen2008,Krauter2011}.

In particular reservoir engineering has been proposed and already used~\cite{Krauter2011} for the generation of steady state nonclassical states, such as linear superposition (Schr\"odinger cat) states~\cite{Poyatos1996,Carvalho2001} or entangled states in microwave cavities~\cite{Pielawa2007,Pielawa2010}. In this case one has the advantage that the desired target state is largely independent of the specific initial states, and at the same time is robust with respect to a large class of decoherence processes. These ideas have been recently extended also to the field of cavity optomechanics for the generation of entangled states of two cavity modes~\cite{Wang2013}, and of two mechanical modes~\cite{Tan2013a}.

Here we apply reservoir engineering for the deterministic generation of robust macroscopic superpositions of coherent states of a mechanical resonator (MR). First proposals for the generation of superposition states exploited the intrinsic nonlinearity of radiation pressure interaction~\cite{Bose1999,Marshall2003} but are hard to realize due to the extremely weak nonlinear coupling. More recent proposals focused on the conditional generation of those linear superposition states~\cite{Paternostro2011,Vanner2013}, exploiting for example the effective measurement of the MR position \emph{squared} in order to generate a superposition of two spatially separated states~\cite{Romero-Isart2011a,Romero-Isart2011,Jacobs2009,Jacobs2011}. These latter schemes do not suffer from weak radiation pressure nonlinearities, but are probabilistic and strongly dependent upon the efficiency of the conditional measurement on the optical mode. As underlined above, the generation of a linear superposition state through reservoir engineering is instead deterministic and extremely robust, because the state is reached asymptotically as a result of a dissipative irreversible evolution, and is less sensitive to the details of the preparation process.

Here we propose to generate a superposition of coherent states of a MR by exploiting the nonlinearity associated with the \emph{quadratic} interaction of the MR with an optical cavity mode, appropriately driven by a bichromatic field (see also Ref.~\cite{Tan2013}). We study the resulting dynamics, determined by the joint action of the engineered reservoir realized by the driven cavity mode and of the standard (and unavoidable) thermal reservoir of the MR. We show that a high-quality superposition state can be generated in a transient time interval, which then decoheres at longer times due to the action of thermal reservoir.  The present scheme is particularly useful for monitoring decoherence processes affecting nanomechanical resonators, similarly to what has been done for cavities~\cite{Deleglise2008} or trapped ions~\cite{Myatt2000}, and could also be useful for testing alternative decoherence models (see Ref.~\cite{Romero-Isart2011} and references therein).

In Sec.~II we describe the properties of the required engineered reservoir. In Sec.~III we show how to engineer such a reservoir by tailoring the optomechanical interaction with a bichromatically driven cavity mode. Sec.~IV describes the resulting dynamics under realistic scenarios, and we verify that a superposition state can be efficiently generated in a transient time interval and that its decoherence can be monitored. Sec.~V is for concluding remarks.

\section{The desired dissipative evolution} \label{sec:proposal}

Let $\rho$ be the reduced density matrix of the MR, and $\rho_\infty=|\psi_\infty\rangle\langle \psi_\infty|$ the target linear superposition state we want to generate in the steady state of the MR, the so-called even Schr\"odinger cat state
\begin{equation}
    |\psi_\infty\rangle =  (|\beta\rangle +|-\beta\rangle)/\mathcal N,
\label{eq:targetstate}
\end{equation}
where $|\beta\rangle $ denotes a coherent state of the MR with complex amplitude $\beta$ and $\mathcal
N=\sqrt{2[1+\exp(-2|\beta|^{2})]}$ is the normalization constant. Reservoir engineering means in the present case tailoring the interaction with the optical cavity mode in order to have an effective reduced dynamics of the MR described by the master equation
\begin{equation}
\frac{\partial}{\partial t}\rho=\mathcal L\rho\,,
\end{equation}
for which $\rho_\infty$ is a fixed point, namely,
\beq
\label{eq:L}
\mathcal L\rho_\infty=0\,.\eeq
A simple solution is to take the Lindbladian ${\mathcal L}$
\begin{eqnarray}
\label{eq:Liouville operators}
  \mathcal L\rho & = & \Gamma {\mathcal D}(C) \rho \\
  {\mathcal D}(C) \rho & = & \left(2C\rho C^{\dagger}-C^{\dagger}C \rho - \rho C^{\dagger}C\right),
\end{eqnarray}
with $\Gamma$ a model-dependent rate, and with the operator $C$ such that $ |\psi_\infty\rangle$ is eigenstate of $C$ with zero eigenvalue. Such a condition is satisfied by
\beq \label{eq: Lindblad operator}
C=(b^2-\beta^{2}),
\eeq
where $b$ is the annihilation operator of the MR, and $\beta $ is just the complex amplitude of the target linear superposition state.

Notice that state $\rho_{\infty}$ is not the unique solution of Eq.~\eqref{eq:L} because any coherent or incoherent superposition of $|\beta\rangle $ and $|-\beta\rangle $ solves Eq.~\eqref{eq:L}. However for our purposes it is sufficient that, at least for a physically realizable class of initial states of the MR, the dissipative evolution asymptotically drives it only to $\rho_{\infty}$ and not to other states of the convex set of states ${\mathcal C_L}$ defined by Eq.~\eqref{eq:L}. In this respect one can profit from an additional symmetry of the Lindbladian of Eq.~(\ref{eq:Liouville operators}), i.e., the fact that it commutes with the parity operator ${\mathcal P}=(-1)^{b^{\dagger} b}$, and therefore ${\mathcal P}$ is conserved as long as the dynamics is driven by $\mathcal L$ only or at least parity non-conserving terms are negligible in the time evolution generator. In such a case, since $|\psi_\infty\rangle$ is the unique pure state of ${\mathcal C_L}$ which is even, that is, eigenstate of ${\mathcal P}$ with eigenvalue $+1$, the asymptotic steady state of the MR will also have parity $+1$. In particular it is possible to see that if the initial state is pure and even, the asymptotic state will be $|\psi_\infty\rangle$. A natural case of this kind is provided by a vacuum initial state $|0\rangle \langle 0|$, which is obtained if the MR is initially cooled to its ground state.

Therefore our goal is to generate the effective reduced dynamics of the MR driven by the above Lindbladian of Eqs.~(\ref{eq:Liouville operators})-(\ref{eq: Lindblad operator}) when the cavity mode is adiabatically eliminated. In practice however, the MR dynamics will be affected not only by the cavity mode ``engineered reservoir'' but also by the standard thermal reservoir. Therefore we have to establish the effect of these undesired latter terms, and to determine if and when they are negligible.

\section{Engineering the dissipative process} \label{sec:Lindblad}

Our starting point is the Hamiltonian of an optomechanical system formed by a driven cavity mode interacting \emph{quadratically} with a MR. Such a quadratic interaction is achieved for example in a membrane-in-the-middle (MIM) setup, when the membrane is placed at a node, or exactly at an avoided crossing point within the cavity \cite{Thompson2008,Sankey2010,Karuza2013}. Alternatively, such a quadratic coupling can be obtained by trapping levitating nanoparticles around an intensity maximum of a cavity mode \cite{Barker2010,Li2011,Gieseler2012,Kiesel2013}. We assume that the cavity is bichromatically driven, that is
\begin{eqnarray}
&&H=\hbar \omega_m b^{\dagger} b +\hbar \omega_c a^{\dagger }a +\hbar g_2 a^{\dagger }a (b+b^{\dagger})^2\\
&&+\mathrm{i}\hbar \left[(E_0 e^{-\mathrm{i}\omega_{\mathrm L} t}+E_1 e^{-\mathrm{i}(\omega_{\mathrm L}+\Omega) t})a^{\dagger }\right. \nonumber \\
&& \left.-(E_0 e^{\mathrm{i}\omega_{\mathrm L} t}+E_1 e^{\mathrm{i}(\omega_{\mathrm L}+\Omega) t})a\right],  \nonumber \label{eq:Ham-optomech}
\end{eqnarray}
where $\omega_m$ is the resonance frequency of the MR, $\omega_c$ the cavity mode frequency, $E_0=\sqrt{2P_{0}\kappa_0 /\hbar \, \omega _{\mathrm L}}$, $E_1=\sqrt{2P_{1}\kappa_0 /\hbar \, (\omega _{\mathrm L}+\Omega)}$, with $\kappa_0$ the cavity decay rate through the input mirror, and $P_0$ and $P_1$ (with $P_0 \gg P_1$) the respective input power at the two driving frequencies. $g_2$ is the quadratic optomechanical coupling rate, which is equal to $g_2 = \Theta (\partial^2 \omega_c/\partial z_0^2)(\hbar/2 m\omega_m)$ in the MIM case, with $\Theta$ the transverse overlap between the mechanical and optical mode at the membrane, and $m$ the membrane mode effective mass \cite{Karuza2013}.

We then move to the frame rotating at the main laser frequency $\omega_{\mathrm L}$, where the system Hamiltonian becomes
\begin{eqnarray}
&&H=\hbar \omega_m b^{\dagger} b +\hbar \Delta_0 a^{\dagger }a +\hbar g_2 a^{\dagger }a (b+b^{\dagger})^2 \label{eq:Ham-optomech2} \\
&&+\mathrm{i}\hbar \left[(E_0+E_1 e^{-\mathrm{i}\Omega t})a^{\dagger }-(E_0 +E_1 e^{\mathrm{i}\Omega t})a\right],  \nonumber
\end{eqnarray}
where $\Delta_0=\omega_c-\omega_{\mathrm L}$ is the cavity mode detuning. The dynamics is however also driven by fluctuation-dissipation processes associated with the coupling of the cavity mode with the optical vacuum field outside the cavity, and of the MR with its thermal reservoir characterized by a temperature $T$ and a mean thermal phonon number $\bar{n}=\left[\exp(\hbar\omega_m/k_B T)-1\right]^{-1}$. In the usual Markovian approximation \cite{Gardiner2000}, we have
that optical dissipation is described by
\begin{equation}\label{eq:optdiss}
    \kappa_{\rm T} {\mathcal D}(a) \rho_{om},
\end{equation}
where $\rho_{om}$ is the density matrix of the total optomechanical system, and $ \kappa_{\rm T}$ is the total cavity decay rate, while mechanical fluctuation-dissipation effects are described by the following terms in the master equation~\cite{Gardiner2000}
\begin{equation}\label{eq:optmech}
    \frac{\gamma_{\rm m}}{2}(\bar{n}+1) {\mathcal D}(b) \rho_{om} + \frac{\gamma_{\rm m}}{2}\bar{n} {\mathcal D}(b^{\dagger}) \rho_{om},
\end{equation}
where $\gamma_m=\omega_m/Q_m$ is the mechanical damping, and $Q_m$ is the mechanical quality factor. Therefore the time evolution of the system is described by the following general master equation
\begin{eqnarray}
\label{eq:meq-gen}
&&\frac{\partial}{\partial t}\rho_{om}=-\frac{\mathrm i}{\hbar}\left[H,\rho_{om}\right]+\kappa_{\rm T} {\mathcal D}(a) \rho_{om} \\
&& +\frac{\gamma_{\rm m}}{2}(\bar{n}+1) {\mathcal D}(b) \rho_{om} + \frac{\gamma_{\rm m}}{2}\bar{n} {\mathcal D}(b^{\dagger}) \rho_{om}, \nonumber
\end{eqnarray}
with $H$ given by Eq.~(\ref{eq:Ham-optomech2}).

The intense driving associated with the laser field at the carrier frequency $\omega_{\mathrm L}$ generates a stationary intracavity state of the cavity mode with large coherent amplitude
\begin{equation}\label{eq:alphas}
    \alpha_s = \frac{E_0}{\kappa_{\rm T}+\mathrm i\Delta_0},
\end{equation}
and it is convenient to look at the dynamics of the quantum fluctuations of the cavity mode, performing the displacement $a = \alpha_s + \delta a$. After some algebra and using Eq.~(\ref{eq:alphas}), the master equation of Eq.~(\ref{eq:meq-gen}) becomes
\begin{eqnarray}
\label{eq:meq-gen2}
&&\frac{\partial}{\partial t}\rho_{om}=-\frac{\mathrm i}{\hbar}\left[H_{\delta},\rho_{om}\right]+\kappa_{\rm T} {\mathcal D}(\delta a) \rho_{om} \\
&& +\frac{\gamma_{\rm m}}{2}(\bar{n}+1) {\mathcal D}(b) \rho_{om} + \frac{\gamma_{\rm m}}{2}\bar{n} {\mathcal D}(b^{\dagger}) \rho_{om}, \nonumber
\end{eqnarray}
with the modified Hamiltonian
\begin{eqnarray}
&&H_{\delta}=\hbar \tilde{\omega}_m b^{\dagger} b +\hbar g_2 |\alpha_s|^2 \left(b^2+b^{\dagger\, 2}\right)+ \hbar \Delta_0 \delta a^{\dagger }\delta a \nonumber \\
&&+\mathrm{i}\hbar \left[E_1 e^{-\mathrm{i}\Omega t}\delta a^{\dagger }-E_1 e^{\mathrm{i}\Omega t}\delta a\right] \\
&& +\hbar g_2 \left(\alpha_s^* \delta a+ \alpha_s \delta a^{\dagger} \right) (b+b^{\dagger})^2+\hbar g_2 \delta a^{\dagger }\delta a (b+b^{\dagger})^2,  \nonumber \label{eq:Ham-optomech-flu}
\end{eqnarray}
where  $ \tilde{\omega}_m = \omega_m + 2 g_2 |\alpha_s|^2$ is the renormalized mechanical frequency.

We now take $\Omega = \Delta_0$, i.e., we assume that the second, less intense beam is exactly resonant with the cavity mode, and move to the interaction picture with respect to
\begin{equation}
H_{0}=\hbar \tilde{\omega}_m b^{\dagger} b + \hbar \Delta_0 \delta a^{\dagger }\delta a.
\end{equation}
Within such a picture, the dissipative terms in the master equation of Eq.~(\ref{eq:meq-gen2}) remain unchanged, while the Hamiltonian becomes
\begin{eqnarray}
&&H_{\delta}^{int}=\hbar g_2 |\alpha_s|^2 \left(b^2 e^{-2\mathrm{i}\tilde{\omega}_m t}+b^{\dagger\, 2}e^{2\mathrm{i}\tilde{\omega}_m t}\right)+\mathrm{i}\hbar \left[E_1 \delta a^{\dagger }-E_1 \delta a\right]  \nonumber  \\
&& +\hbar g_2 \left(\alpha_s^* \delta a e^{-\mathrm{i}\Delta_0 t}+ \alpha_s \delta a^{\dagger}e^{\mathrm{i}\Delta_0 t} \right) (be^{-\mathrm{i}\tilde{\omega}_m t}+b^{\dagger}e^{\mathrm{i}\tilde{\omega}_m t})^2  \nonumber  \\
&&+\hbar g_2 \delta a^{\dagger }\delta a (be^{-\mathrm{i}\tilde{\omega}_m t}+b^{\dagger}e^{\mathrm{i}\tilde{\omega}_m t})^2.
\label{eq:Ham-optomech-flu-int}
\end{eqnarray}
We have made no approximation up to now. We now take the following \emph{resonance condition}, $\Delta_0 = \Omega = 2\tilde{\omega}_m$, which means that the second driving beam is resonant not only with the cavity, but also with the second order sideband of the carrier beam at $\omega_{\mathrm L}$, and make the two following approximations: i) we neglect the last, higher order interaction term $\hbar g_2 \delta a^{\dagger }\delta a (be^{-\mathrm{i}\tilde{\omega}_m t}+b^{\dagger}e^{\mathrm{i}\tilde{\omega}_m t})^2$, which is justified whenever $|\delta a | \ll |\alpha_s|$; ii) we make the rotating wave approximation (RWA) and neglect all the fast-oscillating terms at $\tilde{\omega}_m$ and $2\tilde{\omega}_m$, which is justified in the weak coupling limit $g_2 |\alpha_s| \ll \tilde{\omega}_m$.
The effective interaction picture Hamiltonian of Eq.~(\ref{eq:Ham-optomech-flu-int}) reduces to
\begin{equation}
H_{\rm eff}=\hbar g_2 \alpha_s^* \delta a \left(b^{\dagger \,2}-\mathrm{i}E_1/g_2 \alpha_s^*\right)+ {\rm H.C.},
\label{eq:Ham-optomech-flu-int2}
\end{equation}
and therefore under the conditions specified above, the dynamics of the optomechanical system is described by Eq.~(\ref{eq:meq-gen2}) with $H_{\delta}$ replaced by $H_{\rm eff}$. An analogous effective optomechanical dynamics has been considered in Ref.~\cite{Tan2013}, where however the renormalization of the mechanical frequency $\omega_m \to \tilde{\omega}_m $ has been neglected.

\subsection{Reduced dynamics of the mechanical resonator}

In the bad cavity limit, i.e., when $\kappa_{\rm T} \gg g_2 |\alpha_s|, \gamma_m \bar{n}$, the optical mode fluctuations $\delta a$ can be adiabatically eliminated because they quickly decay and their state always remains close to the vacuum state (see for example Ref.~\cite{Gardiner2000}, pag. 147 and Ref.~\cite{Wiseman1993}). One gets the following final effective master equation for the reduced density matrix of the MR, $\rho$,
\begin{equation}
\label{eq:meq-red}
\frac{\partial}{\partial t}\rho=\Gamma {\mathcal D}(C) \rho +\frac{\gamma_{\rm m}}{2}(\bar{n}+1) {\mathcal D}(b) \rho + \frac{\gamma_{\rm m}}{2}\bar{n} {\mathcal D}(b^{\dagger}) \rho,
\end{equation}
where $\Gamma = g_2^2 |\alpha_s|^2/\kappa_{\rm T}$ and $C$ is just given by Eq.~(\ref{eq: Lindblad operator}), with the superposition state amplitude $\beta^2 = E_1/{\mathrm i}g_2 \alpha_s$. The first term is just the desired term, i.e., the engineered dissipative evolution able to drive the MR asymptotically to the target superposition state $|\psi_{\infty}\rangle$. However, the time evolution of the MR state is also driven by the second and third terms which are due to the coupling with the thermal reservoir at temperature $T$. The latter are ``undesired'' terms, because they drive the MR to a thermal state rather than the desired even Schr\"odinger cat state, and also because they do not conserve the parity. Due to the joint action of these two dissipative evolutions, the asymptotic state achieved by the MR at long times will be different from the desired even superposition state $|\psi_{\infty}\rangle$.  However, if $\Gamma \gg \gamma_m \bar{n}$ so that the effect of the thermal reservoir is negligible, we expect that the target state can be generated at least for a reasonable transient time interval around $\bar{t} \sim 1/\Gamma $. This condition, together with the conditions $|\alpha_s| \gg 1$, and $\kappa_{\rm T}, \tilde{\omega}_m \gg g_2 |\alpha_s|$ which are needed for deriving Eq.~(\ref{eq:meq-red}), represent the parameter conditions for realizing the robust generation of a superposition state of the MR.

\section{Results}

Let us now we verify if and when the proposal is implementable in a state-of-the-art optomechanical setup and a nanomechanical resonator can be prepared with high fidelity, at least for a long-lived transient, in the macroscopic superposition state $|\psi_{\infty}\rangle$. We consider parameter values achievable in state-of-the-art MIM setups~\cite{Thompson2008,Sankey2010,Karuza2013,Wilson2009,Purdy2012,Flowers-Jacobs2012,Karuza2012,Purdy2013}. For the mechanical resonator we take $\omega_m = 10$ MHz, $ \gamma_m = 0.1$ Hz (implying $Q_m = 10^8$), $m= 1$ ng and we can take $\partial^2 \omega_c/\partial z_0^2= 2\pi \times 20$ GHz/nm$^2$ \cite{Flowers-Jacobs2012}, yielding $g_2 \simeq 5$ Hz. We then take a laser with frequency $\omega_{\rm L} = 1.77 \times 10^{15}$ Hz (corresponding to a wavelength $\lambda = 1064$ nm) and input power $P_0 = 40$ mW. We also choose a cavity with total decay rate $\kappa_{\rm T} = 10^5$ Hz and with decay rate through the input mirror $\kappa_0 \sim \kappa_{\rm T}/2$, yielding $E_0 \sim 1.5 \times 10^{11}$ Hz. The corresponding value of the intracavity amplitude from Eq.~(\ref{eq:alphas}) is $|\alpha_s| \sim 3.45 \times 10^3$. As a consequence $g_2 |\alpha_s| \sim 1.7 \times 10^4$ Hz, which therefore agrees with the assumptions made. Moreover we have an effective decay rate $\Gamma = g_2^2 |\alpha_s|^2/\kappa_{\rm T} \sim 2.98$ kHz which is reasonably larger than the thermal decay rate $\gamma_m \bar{n}$ as long as $\bar{n} \lesssim 100$. Even though nontrivial, this latter condition is achievable in current optomechanical experiments because cryogenic environments at temperatures $T \simeq 10$ mK are feasible and, with the chosen value $\omega_m = 10 $ MHz, this corresponds just to $\bar{n}\simeq 100$.

The amplitude $\beta$ of the target state is determined by $E_1$ and therefore by the input power $P_1$. Assuming $P_1 \sim 1$ pW, one gets $ E_1 \sim 10^6 $ Hz and therefore $|\beta| \sim 23.6$, which corresponds to a quite macroscopic superposition state; here however, in order to verify numerically the proposal in a not too large operational Hilbert space, we have taken $P_1 \sim 0.01$ pW, yielding $ E_1 \sim 10^5 $ Hz and therefore $|\beta| \sim 2.36$.

\subsection{Cat state generation starting from the mechanical ground state}

As discussed above, we expect to generate a long-lived transient even Schr\"odinger cat state of the MR when $\Gamma \gg \gamma_m \bar{n}$ and if we start from the mechanical ground state, which is pure and even. Since in the considered scenario it is very hard to go below $\bar{n}\sim 100$ with cryogenic techniques only, this initial state could be achieved, at least in principle, by first laser cooling the MR to its ground state, i.e., by first considering a \emph{linear} optomechanical interaction with a cavity mode and driving it on its first red sideband \cite{Marquardt2007,Wilson-Rae2007,Genes2008,Chan2011,Verhagen2012}. Then, one should switch to the quadratic optomechanical interaction (either by displacing the membrane or by driving a different appropriate cavity mode) soon after ground state cooling is attained.

\begin{figure*}[tb]
\subfigure[]{\includegraphics[scale=.33]{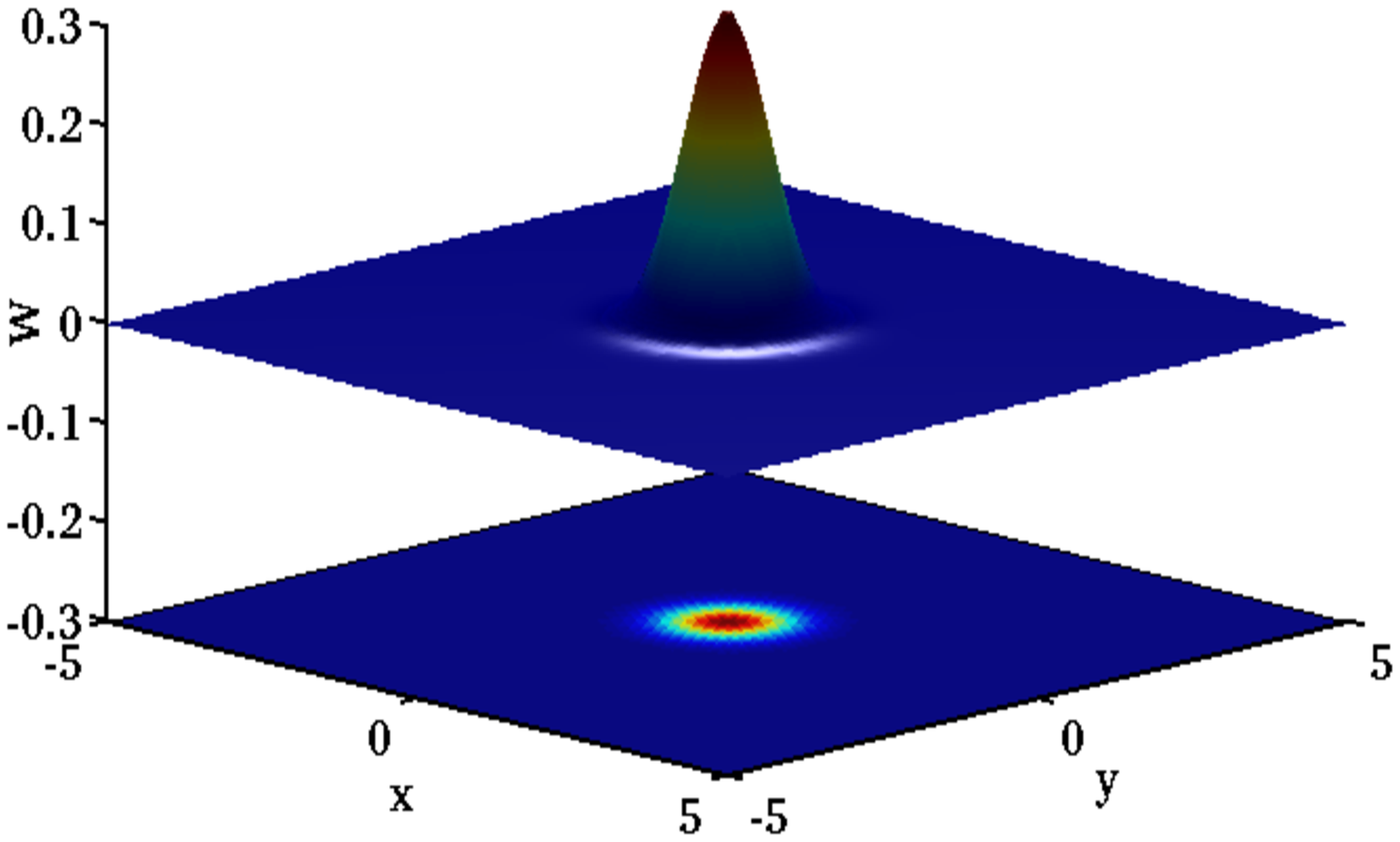}}
\subfigure[]{\includegraphics[scale=.33]{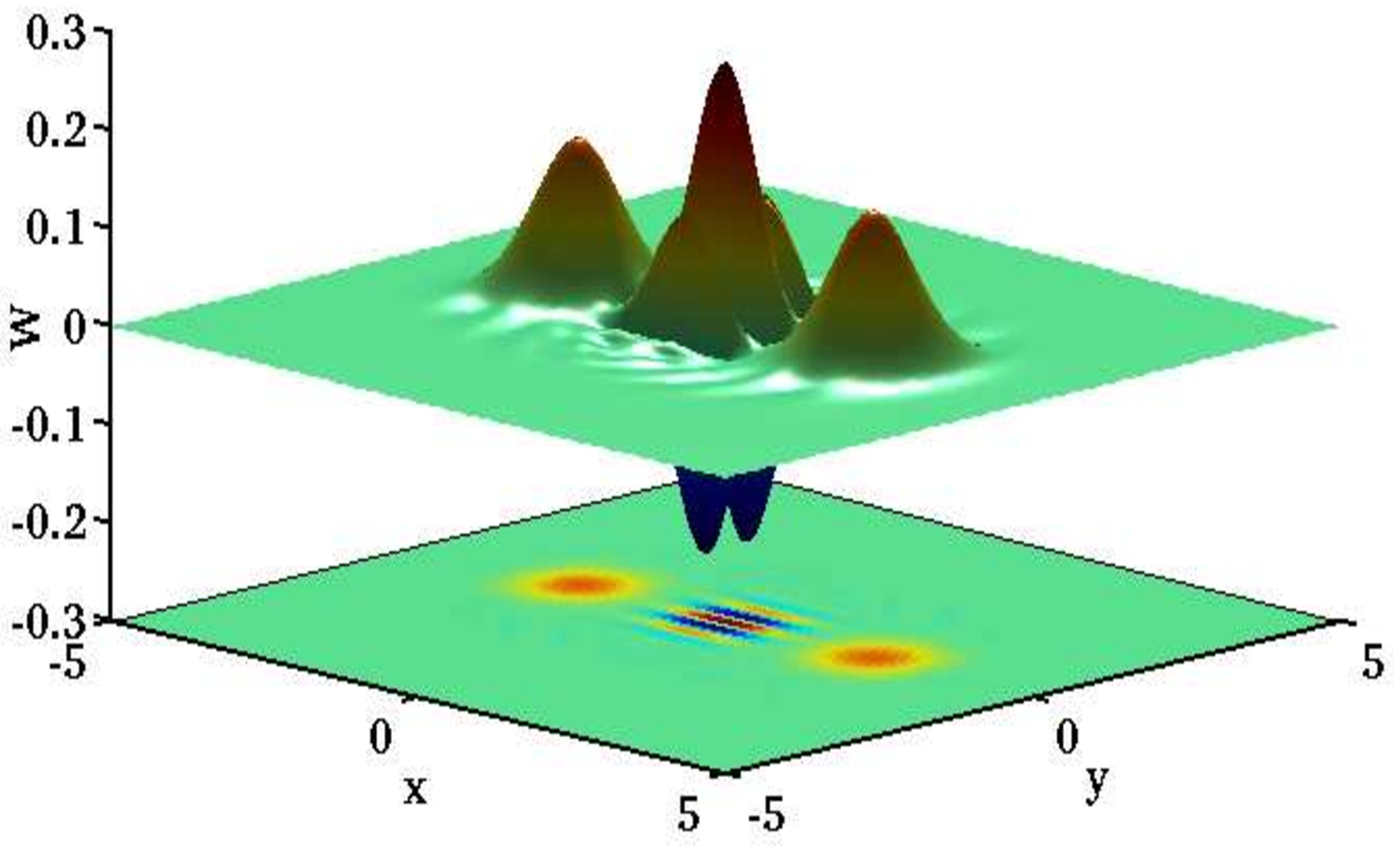}}
\subfigure[]{\includegraphics[scale=.33]{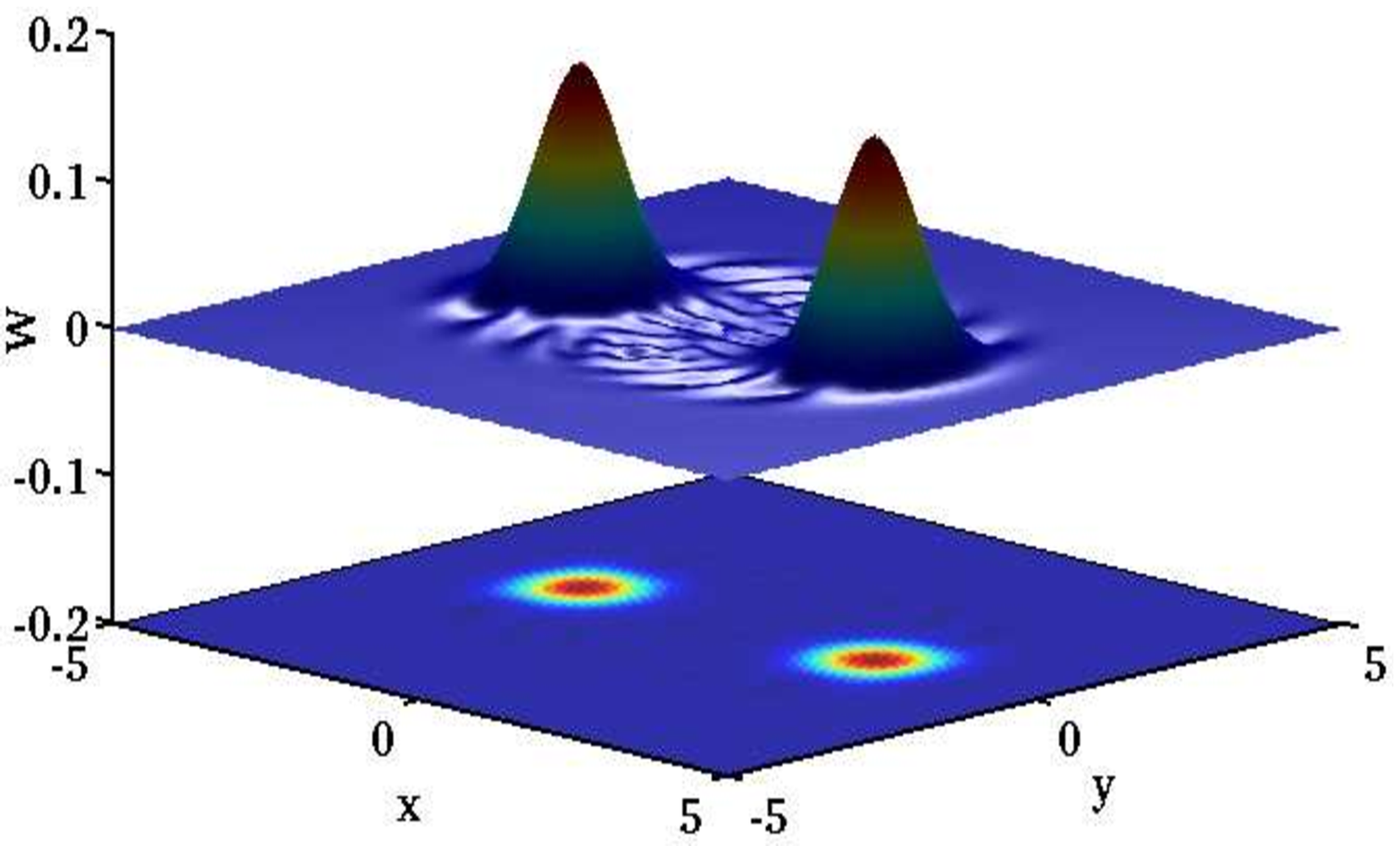}}
\caption{Time evolution of the Wigner function of the reduced state of the MR starting from an initial factorized state in which both the optical cavity fluctuations and the mechanical mode are in their ground state. The set of parameters is given in the text, and $\bar{n}=100$. (a) Wigner function of the initial state of the MR; (b) Wigner function of the MR state at at time $t=0.71/\Gamma=2.39\times 10^{-4}$ s;  (c) Wigner function of the MR state at time $t=100/\Gamma=0.03355$ s. The even superposition state is successfully generated in a short time of the order of $1/\Gamma$, and it slowly loses its nonclassical interference fringes at a longer timescale, of the order of $\left[\gamma_m \left( 2\bar{n}+1\right)\right]^{-1}$.}\label{fig:wig1}
\end{figure*}

We have numerically solved the time evolution of the optomechanical system density matrix $\rho_{om}$ as described by Eq.~(\ref{eq:meq-gen2}) with the Hamiltonian of Eq.~(\ref{eq:Ham-optomech-flu-int2}), starting from the mechanical ground state and the vacuum state for the cavity mode fluctuations. Plots of the Wigner representation of the reduced state $\rho$ of the MR at different times are shown in Fig.~\ref{fig:wig1}, which refers to the set of parameters described above, and $\bar{n}=100$. These plots confirm our expectations and that the state $|\psi_{\infty}\rangle$ with $|\beta| \sim 2.36$ is generated in the transient regime $t \sim 1/\Gamma$ due to the appropriate bichromatic driving and the quadratic optomechanical interaction. The superposition state then decoheres on a time scale governed by $\gamma_m \left( 2\bar{n}+1\right)$. These results are consistent with those of Ref.~\cite{Tan2013} which also studies the generation of a cat state of a MR starting from the ground state in a bichromatically driven quadratic optomechanical system by means of the Wigner function of the reduced MR state.

\begin{figure}[tb]
\includegraphics[scale=.65]{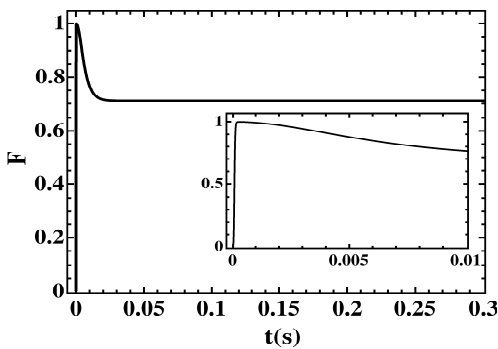}
\caption{Plot of fidelity $ {\mathcal F}(t)={\mathcal F}(\rho_{\infty},\rho(t))$ as a function of time. The inset shows the behavior at short times. Parameters are those given in the text and coinciding with those of Fig.~\protect\ref{fig:wig1}. }\label{fig:fid1}
\end{figure}

This qualitative analysis based on the Wigner function is confirmed by a quantitative analysis based on the time evolution of the fidelity of the state with respect to the target state $|\psi_{\infty}\rangle$. Rather than the more common Uhlmann fidelity~\cite{Uhlmann1976,Jozsa1994}, in order to simplify the numerical calculation, here we use the Hilbert-Schmidt fidelity introduced in Ref.~\cite{Wang2008}
\begin{equation}\label{eq:fid}
    {\mathcal F}(\rho_0,\rho_1) = \frac{\left|{\rm Tr}\left\{\rho_0 \rho_1\right\}\right|}{\sqrt{{\rm Tr}\left\{\rho_0^2\right\}{\rm Tr}\left\{\rho_1^2\right\}}}.
\end{equation}
When $\rho_0$ is pure, this fidelity coincides with the probability of finding the state $\rho_0$ being in $\rho_1$, divided by the square root of the purity $\sqrt{{\rm Tr}\left\{\rho_1^2\right\}}$. In Fig.~\ref{fig:fid1} we plot the time evolution of $  {\mathcal F}(t)={\mathcal F}(\rho_{\infty},\rho(t))$ corresponding to the same parameter condition of Fig.~\ref{fig:wig1}. The fidelity reaches a maximum ${\mathcal F} \simeq 0.9992$ at $t \simeq 1/\Gamma$ when an almost perfect cat state is generated, which then decays so that ${\mathcal F} \simeq 1/\sqrt{2} \simeq 0.7$.

\subsection{Cat state generation after two-phonon cooling}

Fast switching from the linear optomechanical interaction needed for cooling to the mechanical ground state to the quadratic optomechanical interaction necessary for generating the even cat state is quite challenging in practical experimental situations. However, one could exploit the quadratic interaction also for pre-cooling the MR and avoid using a different cavity mode and different driving field. To be more specific one could use the same interaction Hamiltonian and parameter conditions described in the previous section and consider the special case $E_1 = \beta = 0$, i.e., with the weak resonant field turned off. In this case, the engineered interaction with the cavity mode induces a two-phonon cooling process driving the MR to its ground state. The joint dynamics in the presence of nonlinear two-phonon damping and standard decay to the thermal equilibrium with $\bar{n}$ thermal phonons has been already studied in Ref.~\cite{Nunnenkamp2010}, where it is shown that in the limit $\Gamma \gg \gamma_m \bar{n}$ we are considering, cooling is good even though not perfect, being the MR steady state a mixture of the zero and one phonon state, with probabilities $\rho_{11}(\infty)= n_{\rm eff}=(4+1/\bar{n})^{-1}$ and $\rho_{00}(\infty)= 1-\rho_{11}(\infty)$. Therefore a feasible cat state generation protocol is to first cool the MR with the two-phonon cooling process with $E_1=0$, and then switch on the weak resonant field with $E_1\neq 0$ for generating the even cat state as discussed above. We now see that despite the initial approximate $25\%$ probability of being in the odd one phonon state, the cat state generation process is still quite efficient, showing that such a robust macroscopic superposition can be generated in achievable quadratic optomechanical setups.

\begin{figure*}[tb]
\subfigure[]{\includegraphics[scale=.33]{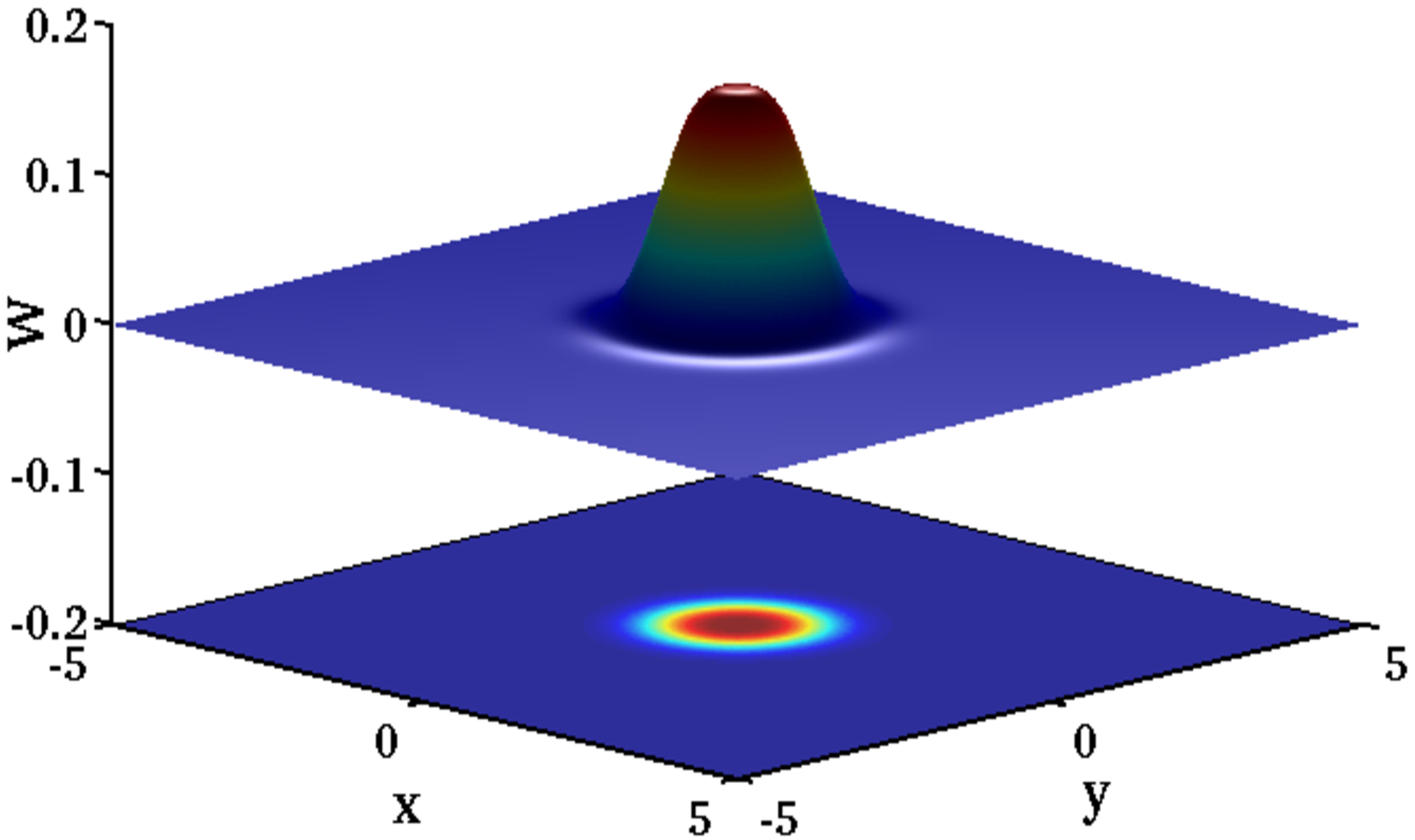}}
\subfigure[]{\includegraphics[scale=.33]{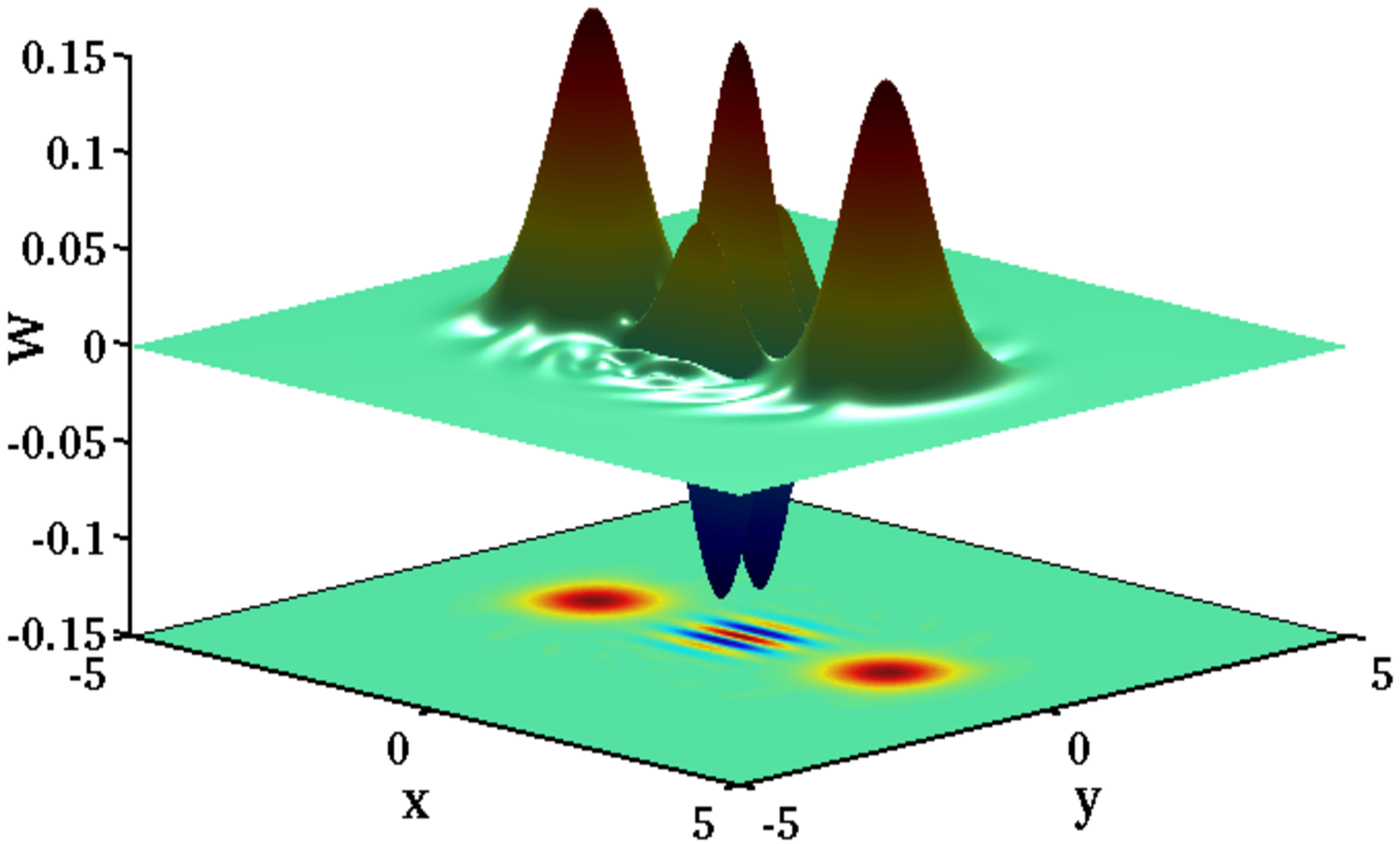}}
\subfigure[]{\includegraphics[scale=.33]{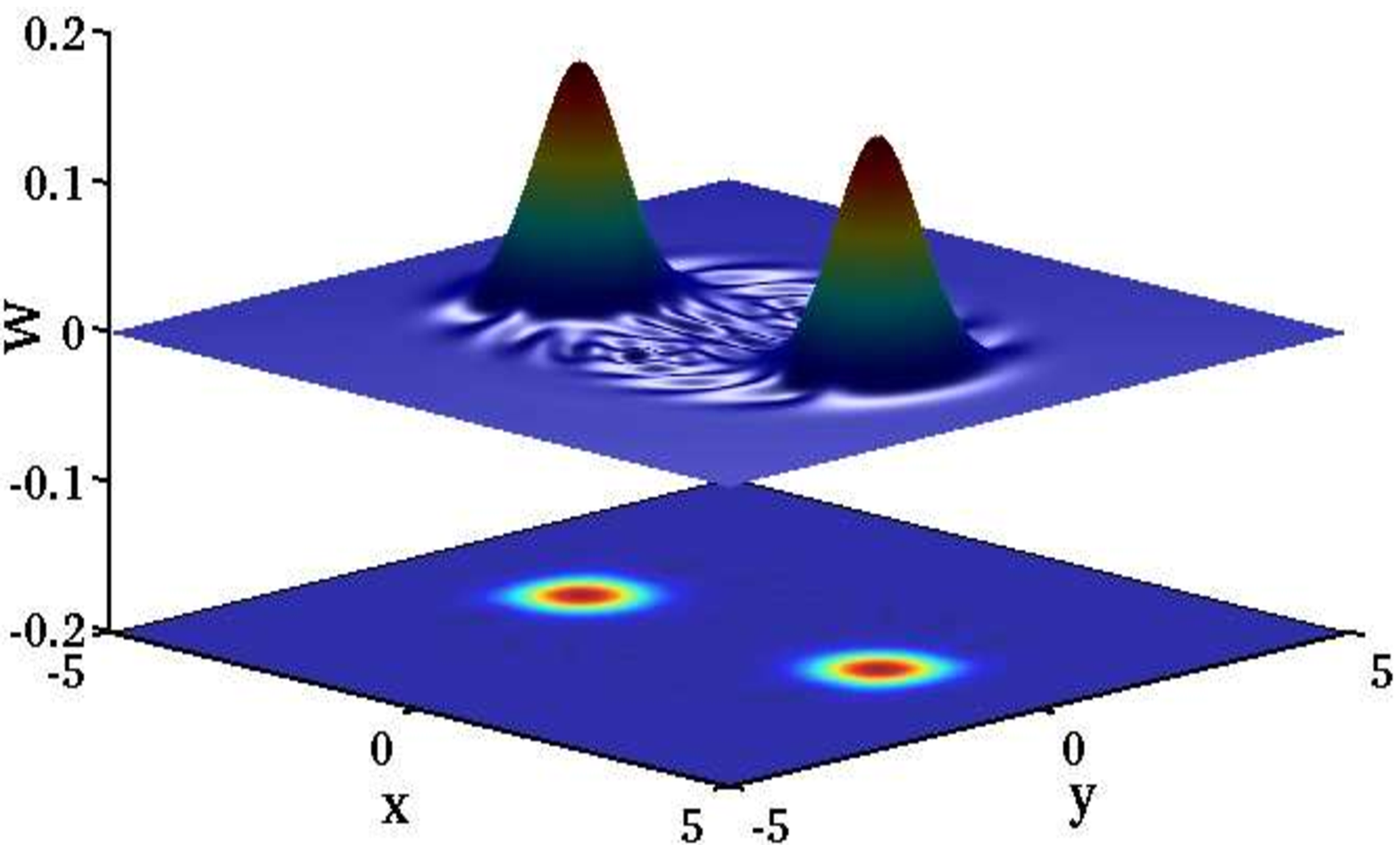}}
\caption{Time evolution of the Wigner function of the reduced state of the MR with mean phonon number $\bar{n}(0)=10$. (a) Wigner function of the initial state of the MR at time t=0; (b) Wigner function of the MR state at time $t=1/\Gamma=3.3547\times 10^{-4}$ s; (c) Wigner function of the MR state at time $t=1000/\Gamma=0.3355$ s. The other parameters are given in the text and coincide with those of Fig.~\protect\ref{fig:fid1}.} \label{fig:wig2}
\end{figure*}

\begin{figure*}[tb]
\subfigure[]{\includegraphics[scale=.33]{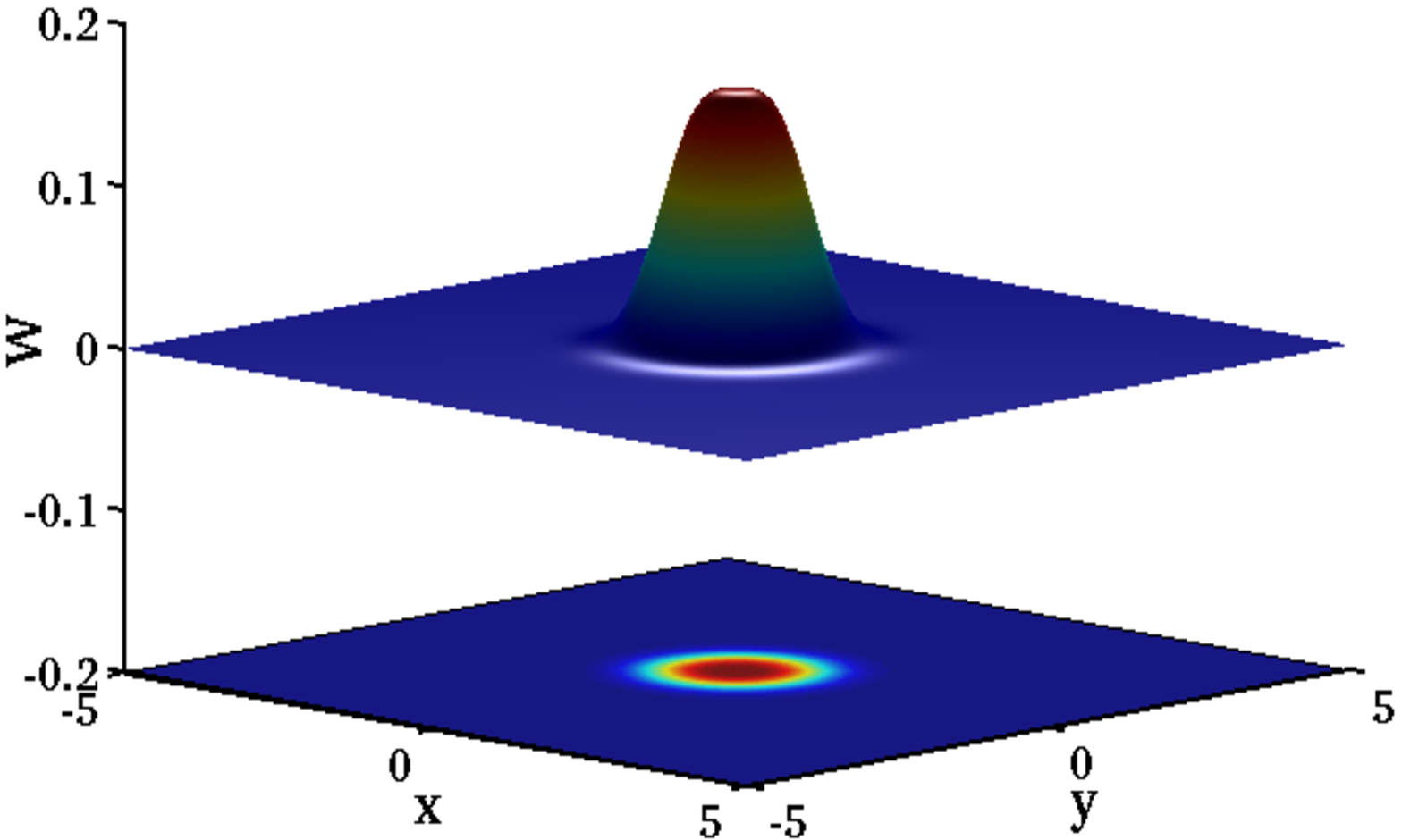}}
\subfigure[]{\includegraphics[scale=.33]{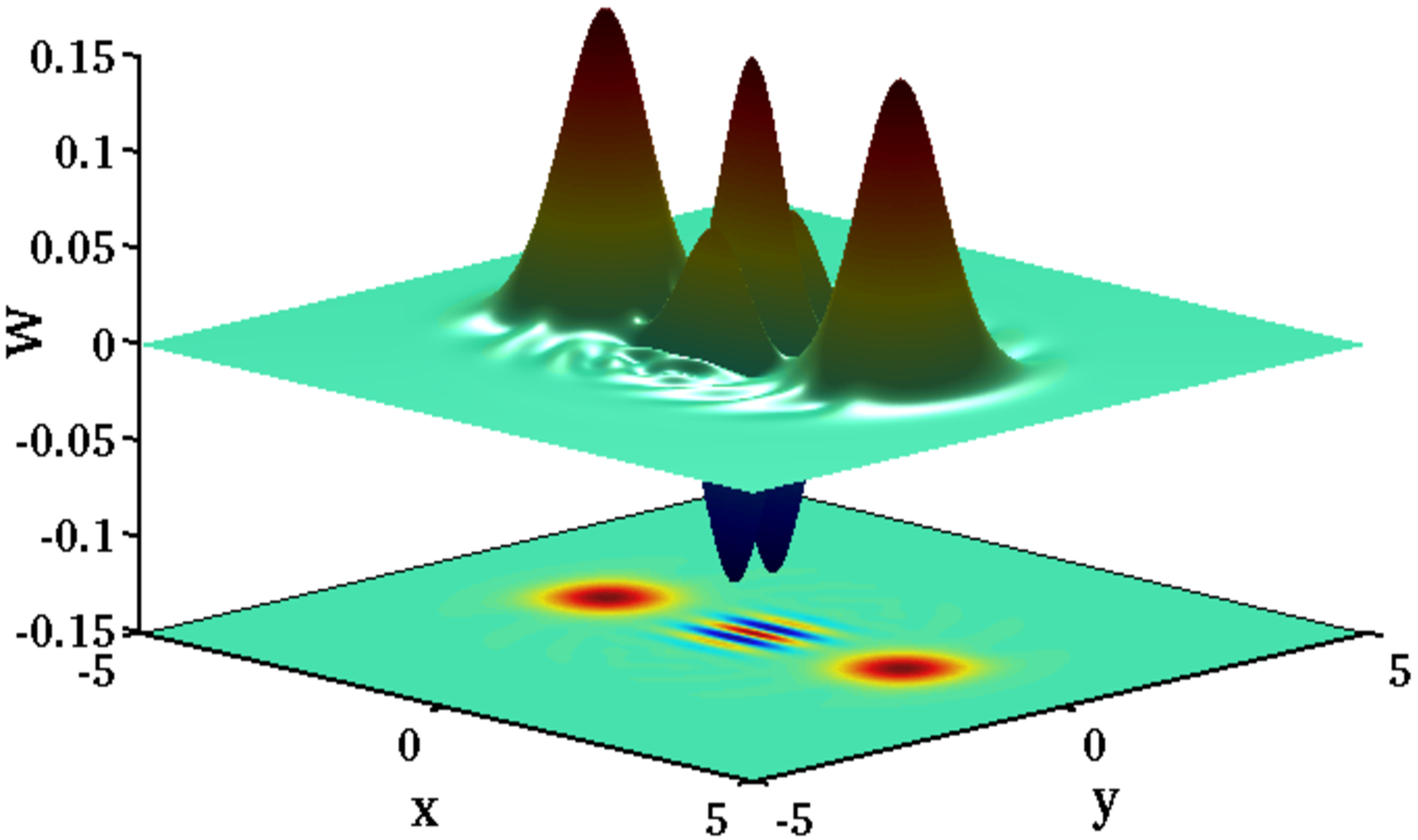}}
\subfigure[]{\includegraphics[scale=.33]{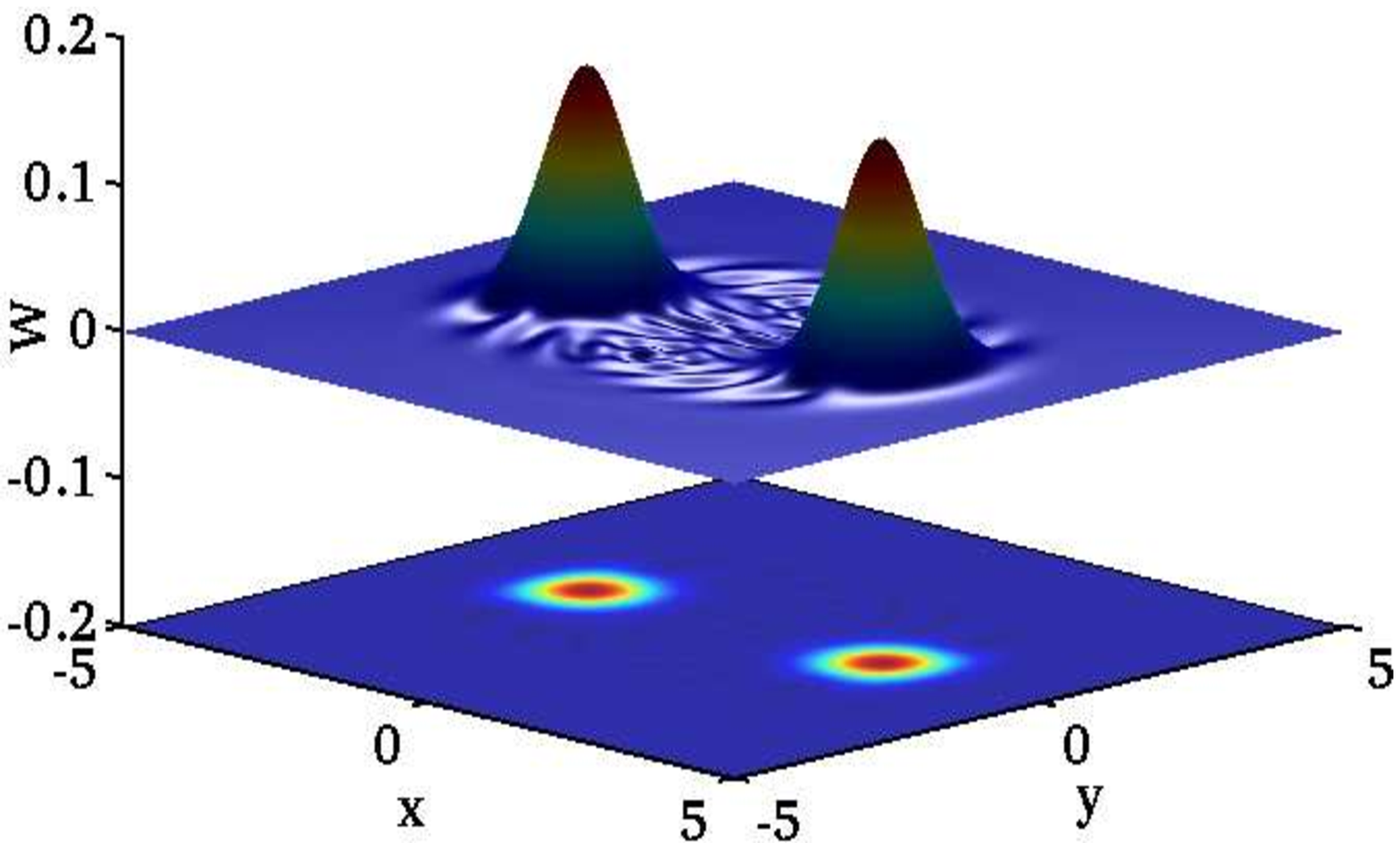}}
\caption{Time evolution of the Wigner function of the reduced state of the MR with mean phonon number $\bar{n}(0)=100$. (a) Wigner function of the initial state of the MR at time t=0; (b) Wigner function of the MR state at time $t=1/\Gamma=3.3547\times 10^{-4}$ s; (c) Wigner function of the MR state at time $t=100/\Gamma=0.03355$ s. The other parameters are given in the text and coincide with those of Fig.~\protect\ref{fig:fid1}.} \label{fig:wig3}
\end{figure*}

\begin{figure*}[tb]
\subfigure[]{\includegraphics[scale=.55]{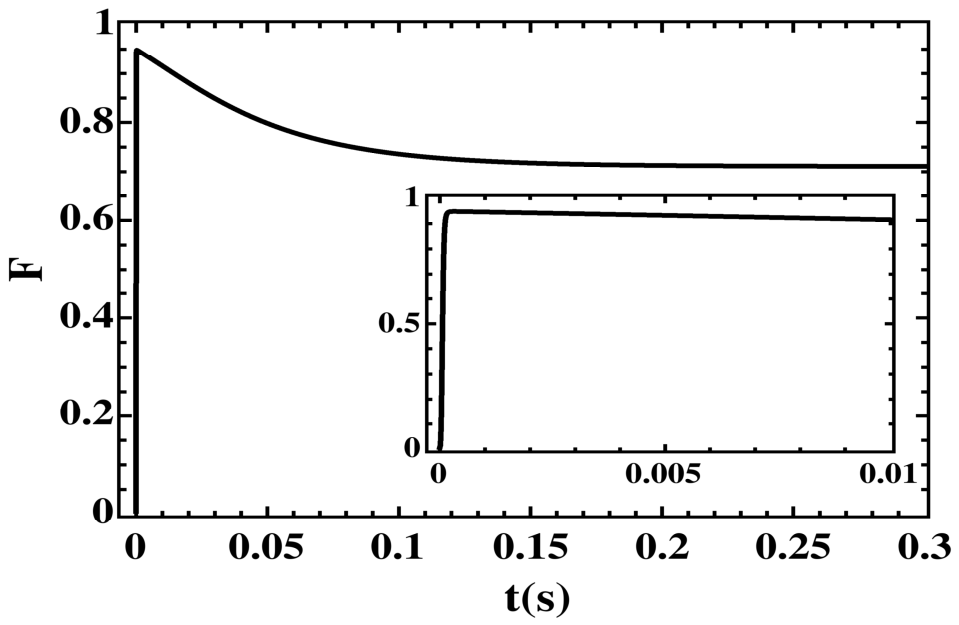}}
\subfigure[]{\includegraphics[scale=.55]{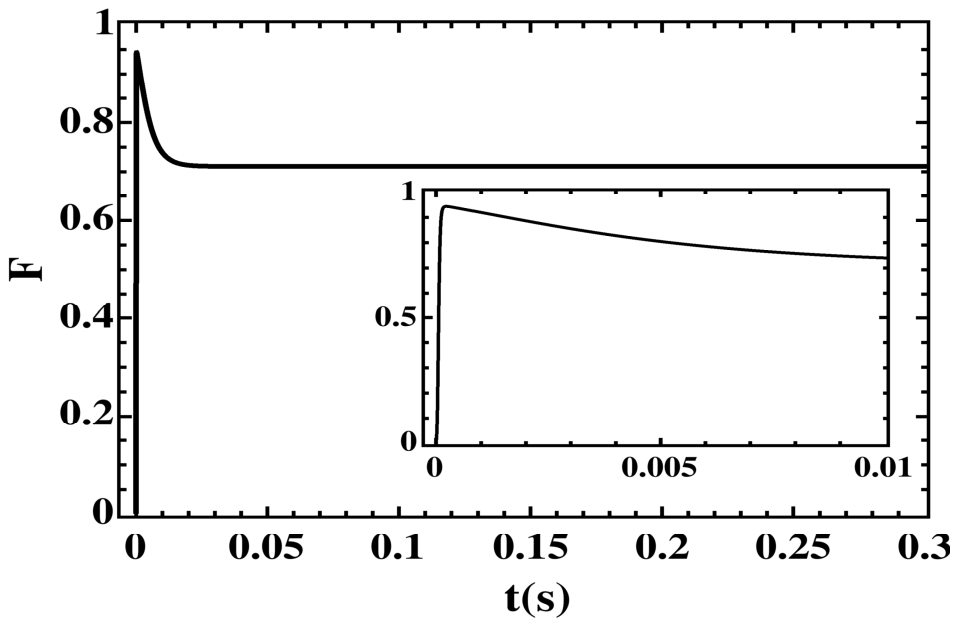}}
\caption{Plot of fidelity $ {\mathcal F}(t)={\mathcal F}(\rho_{\infty},\rho(t))$ as a function of time for (a) $\bar{n}=10$ and (b) $\bar{n}=100$. The other parameters are given in the text and coincide with those of Fig.~\protect\ref{fig:fid1}. The insets show the behavior at short times. } \label{fig:fid2}
\end{figure*}

We have in fact numerically solved the master equation for the optomechanical system density matrix $\rho_{om}$ of Eq.~(\ref{eq:meq-gen2}) with the Hamiltonian of Eq.~(\ref{eq:Ham-optomech-flu-int2}), now taking as initial state the vacuum state for the cavity mode fluctuations and the above mixture of the zero and one phonon state for the MR, using the same set of parameters of the previous subsection (we have verified that with this set of parameters one actually cools the MR to this mixture of states). Plots of the Wigner representation of the reduced state $\rho$ of the MR at different times are shown in Fig.~\ref{fig:wig2}, which refer to $\bar{n}=10$ and in Fig.~\ref{fig:wig3}, which refers to $\bar{n}=100$. In both cases the target cat state is generated with high fidelity at $t \sim 1/\Gamma$, despite the residual excitation in the one-phonon state. This is confirmed by time evolution of the fidelity of the state with respect to the target state $|\psi_{\infty}\rangle$, which is shown in Fig.~\ref{fig:fid2} for $\bar{n}=10$ (a) and $\bar{n}=100$ (b). The fidelity reaches a maximum ${\mathcal F} \simeq 0.94$ at $t \simeq 1/\Gamma$ which does not depend upon $\bar{n}$ and then decays to ${\mathcal F} \simeq 1/\sqrt{2} \simeq 0.7$. The superposition state decoheres to a mixture of two Gaussian states on a time scale governed by the thermal decoherence rate given by $\gamma_{\rm dec}=2\gamma_m |\beta|^2\left( 2\bar{n}+1\right)$~\cite{Kennedy1988,Kim1992,Brune1996,Deleglise2008}.


\subsection{Approximate description of the progressive decoherence of the generated superposition state}

The above analysis shows that the combined action of the engineered reservoir term with rate $\Gamma$ and the thermal reservoir terms with rate $\gamma_m \bar{n}$, when $\Gamma \gg \gamma_m \bar{n}$, generates a superposition state at time $t \simeq 1/\Gamma$ which then decoheres with decoherence rate $2\gamma_m |\beta|^2\left(2\bar{n}+1\right)$.  In particular, Figs.~\ref{fig:wig1}-\ref{fig:wig3} suggest that the MR decoheres to an asymptotic state given by the mixture of the two coherent states $|\pm \beta\rangle \langle\pm \beta |$, with $\beta= \sqrt{E_1/ig_2 \alpha_s}$ just the amplitude of the target superposition state. To state it in other words, the combined action of the engineered and ``natural'' reservoirs tends to stabilize such a mixture of coherent states emerging after the decoherence process. Taking into account the well-established theory of decoherence of superposition of two coherent state in the presence of a thermal reservoir~\cite{Kennedy1988,Kim1992,Brune1996,Deleglise2008}, one is led to approximate the time evolution of the reduced MR state after a transient time $t \geq t_0 \simeq 1/\Gamma$ with the following expression
\begin{eqnarray}\label{eq:rhoapp}
&&\rho_{\rm app}(t>t_0)= {\cal N}(t-t_0)^{-1}\left\{|\beta\rangle \langle \beta|+|-\beta\rangle \langle -\beta| \right. \\
&&\left.+e^{-\left(1+2\bar{n}\right)\gamma_m \left(t-t_0\right)}\left[|\beta\rangle \langle -\beta|+|-\beta\rangle \langle \beta|\right]\right\}, \nonumber
\end{eqnarray}
with ${\cal N}(t)=2\left[1+e^{-2|\beta|^2} e^{-\left(1+2\bar{n}\right)\gamma_m t}\right]$, describing a decohering cat state, which decoheres to its corresponding mixture just at the rate $2\gamma_m |\beta|^2\left(2\bar{n}+1\right)$.

\begin{figure*}[tb]
\subfigure[]{\includegraphics[scale=.45]{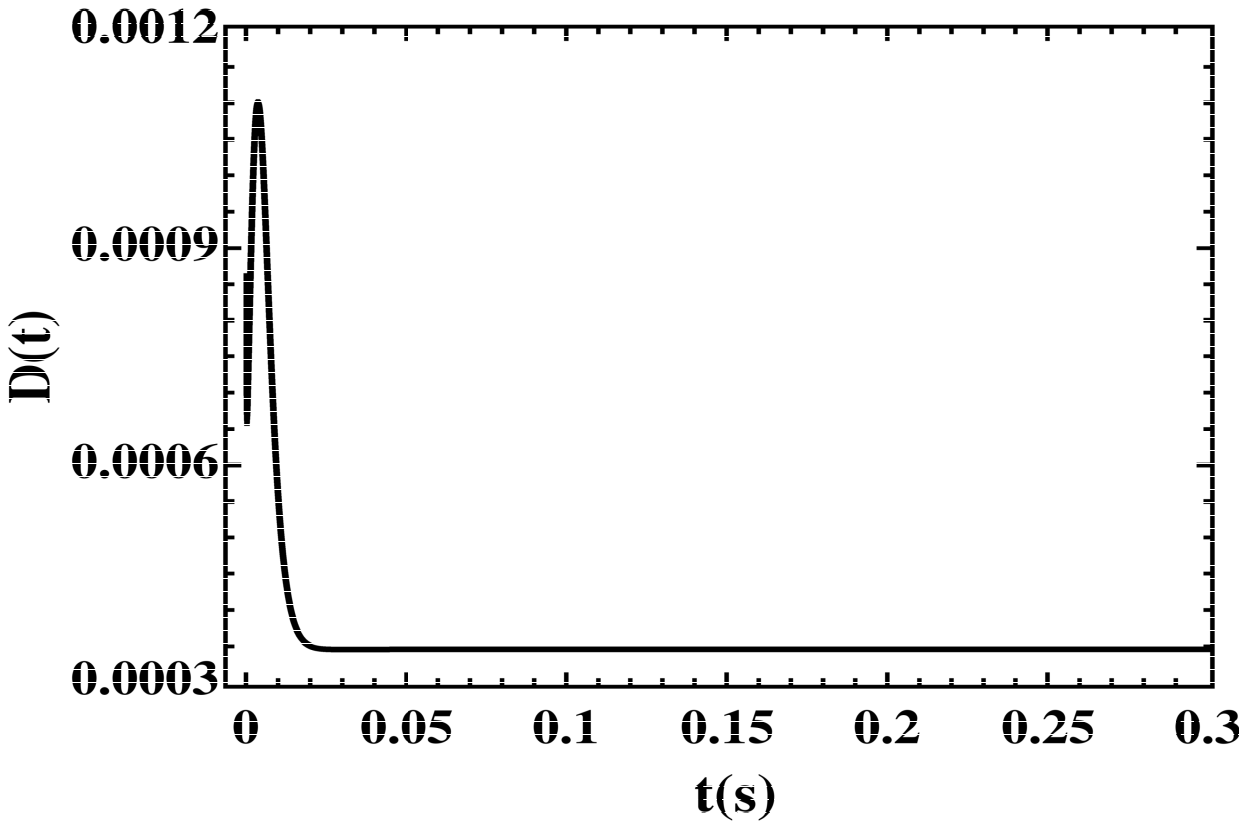}}
\subfigure[]{\includegraphics[scale=.45]{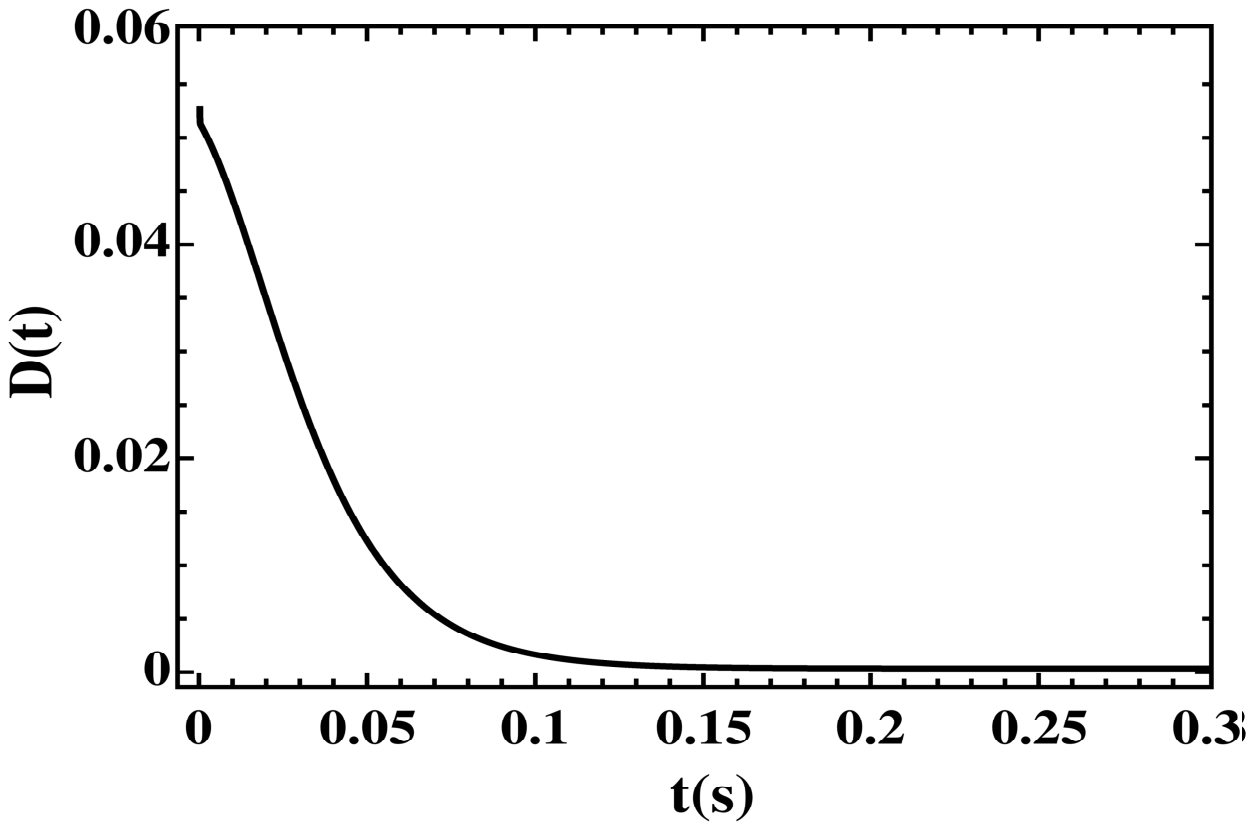}}
\subfigure[]{\includegraphics[scale=.45]{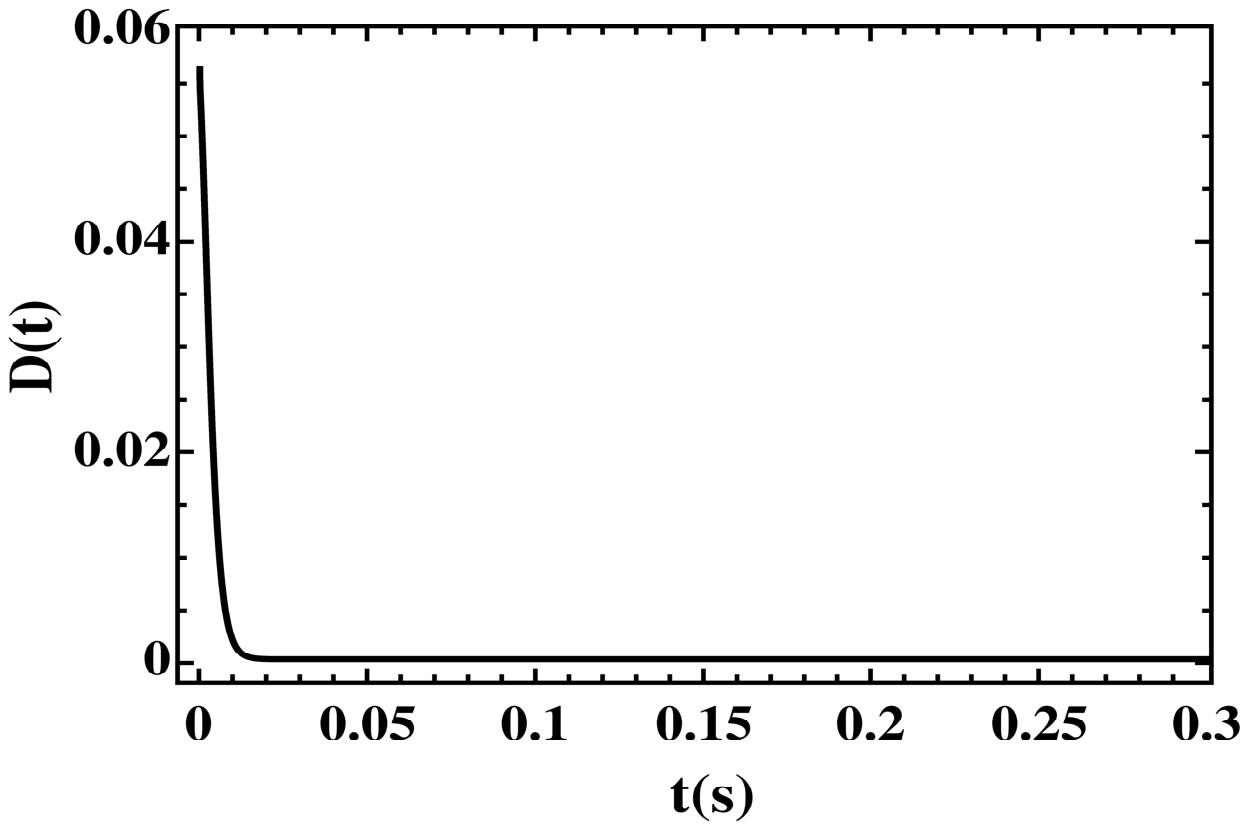}}
\caption{Plot of the distance between the actual solution of the master equation $\rho(t)$ and the approximate MR state of Eq.~(\protect\ref{eq:rhoapp}) as a function of time for (a)$\bar{n}=100$ and initial ground state for the MR; (b) $\bar{n}=10$ and the mixture of zero and one phonon state as initial state of the MR; (c) $\bar{n}=100$ and the mixture of zero and one phonon state as initial state of the MR. The other parameters are given in the text and coincide with those of Fig.~\protect\ref{fig:fid1}. }\label{fig:infid}
\end{figure*}

We can check the validity of this approximate description at $t > t_0$ by using again the Hilbert-Schmidt fidelity of Eq.~(\ref{eq:fid}) for measuring the overlap between the actual reduced MR state $\rho(t)$ given by the solution of the master equation of Eq.~(\ref{eq:meq-gen2}) and the approximate solution $\rho_{\rm app}$ of Eq.~(\ref{eq:rhoapp}). In Fig.~\ref{fig:infid} we plot the ``distance'' between the two states, $D(t)=1-{\mathcal F}\left(\rho(t),\rho_{\rm app}(t)\right)$ for the same set of parameters of Fig.~\ref{fig:fid1}, and we find a very good agreement for the proposed solution. Therefore the generated Schr\"odinger cat state can be used for verifying experimentally the decoherence processes affecting the nanomechanical resonator and eventually testing alternative decoherence models, as suggested in Ref.~\cite{Romero-Isart2011}.

\subsection{Cat state decoherence as decay of non-Gaussianity}

The decoherence process affecting the MR state can also be described as a dynamical ``Gaussification'' process in which the non-Gaussian even cat state generated at short times by the engineered two-phonon reservoir becomes at long times a convex mixture of Gaussian state, i.e., the equal-weight incoherent superposition of the two coherent states $|\pm \beta\rangle$. This suggests an alternative quantitative description of the above loss of quantum coherence caused by the interplay between the engineered and natural reservoir in terms of a measure of \emph{quantum non-Gaussianity} recently proposed in Refs.~\cite{Genoni2013,Palma2013}. 

A state is quantum non-Gaussian if it cannot be written as a convex sum of Gaussian states, and a simple sufficient condition for non-Gaussianity can be given in terms of the value of the Wigner function of the state at the phase space origin $W[\rho](0)$~\cite{Genoni2013}: $\rho$ is quantum non-Gaussian if $W[\rho](0) < (2/\pi)\exp[-2\langle n\rangle(\langle n\rangle +1)]$, where $\langle n \rangle = {\rm Tr}\left\{\rho b^{\dagger}b\right\}$ is the mean number of excitations. However this condition does not detect many quantum non-Gaussian states (for example even cat states) and a more efficient condition for detecting quantum non-Gaussian states has been derived in Ref.~\cite{Palma2013}: $\rho$ is quantum non-Gaussian if there is a Gaussian map ${\mathcal E}$ such that 
\begin{equation}
\label{eq:ng}
NG = W[{\mathcal E}(\rho)](0) - \frac{2}{\pi}\exp[-2\langle n_{\mathcal E}\rangle(\langle n_{\mathcal E}\rangle +1)<0,
\end{equation}
where ${\mathcal E}(\rho)$ is the state transformed by the Gaussian map and $ n_{\mathcal E}$ is the mean excitation number of the transformed state.

\begin{figure*}[tb]
\subfigure[]{\includegraphics[scale=.45]{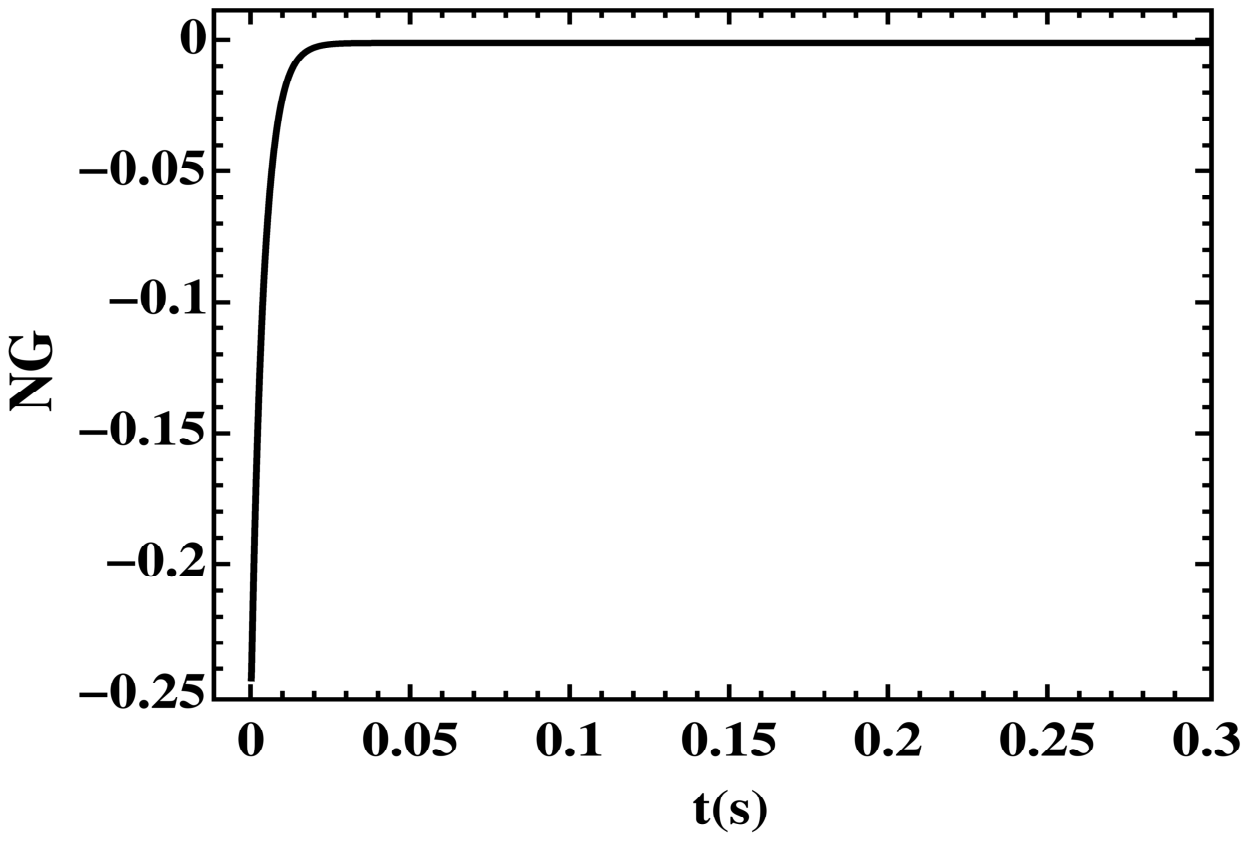}}
\subfigure[]{\includegraphics[scale=.45]{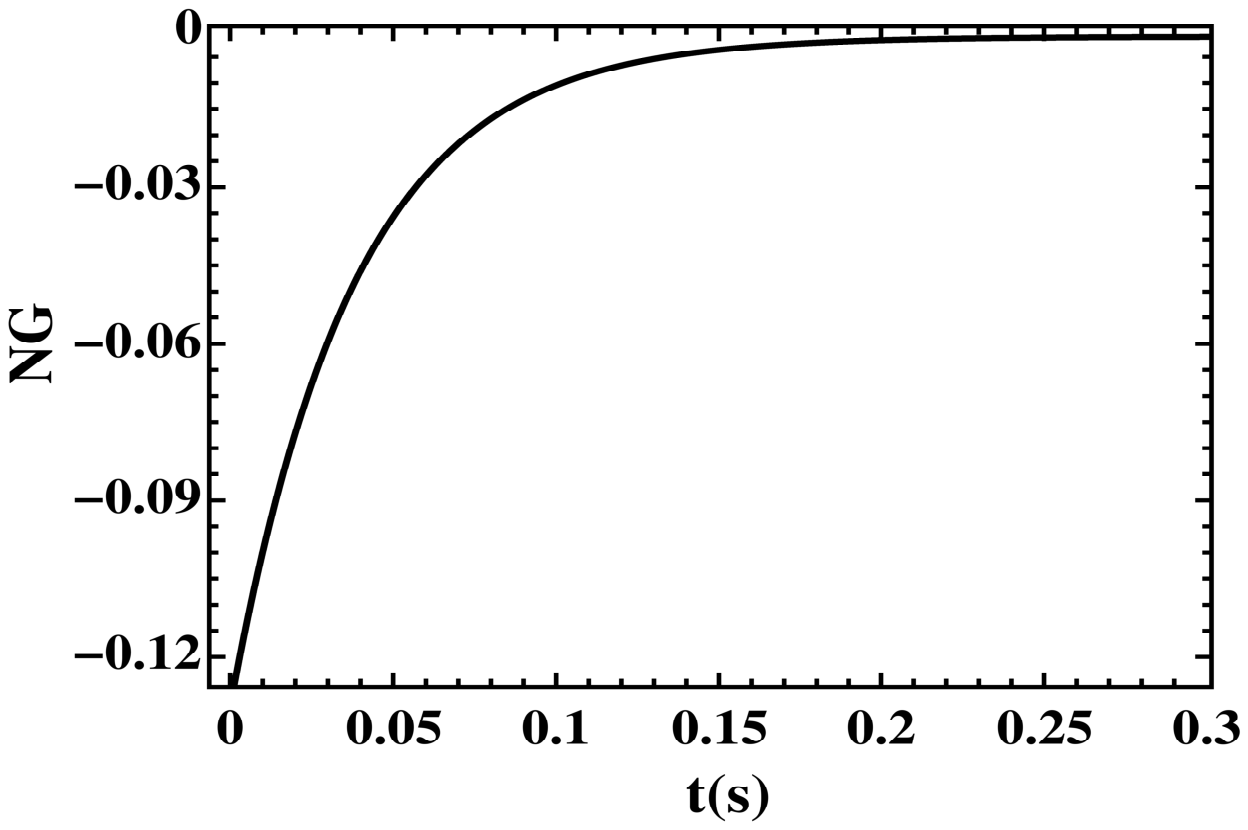}}
\subfigure[]{\includegraphics[scale=.45]{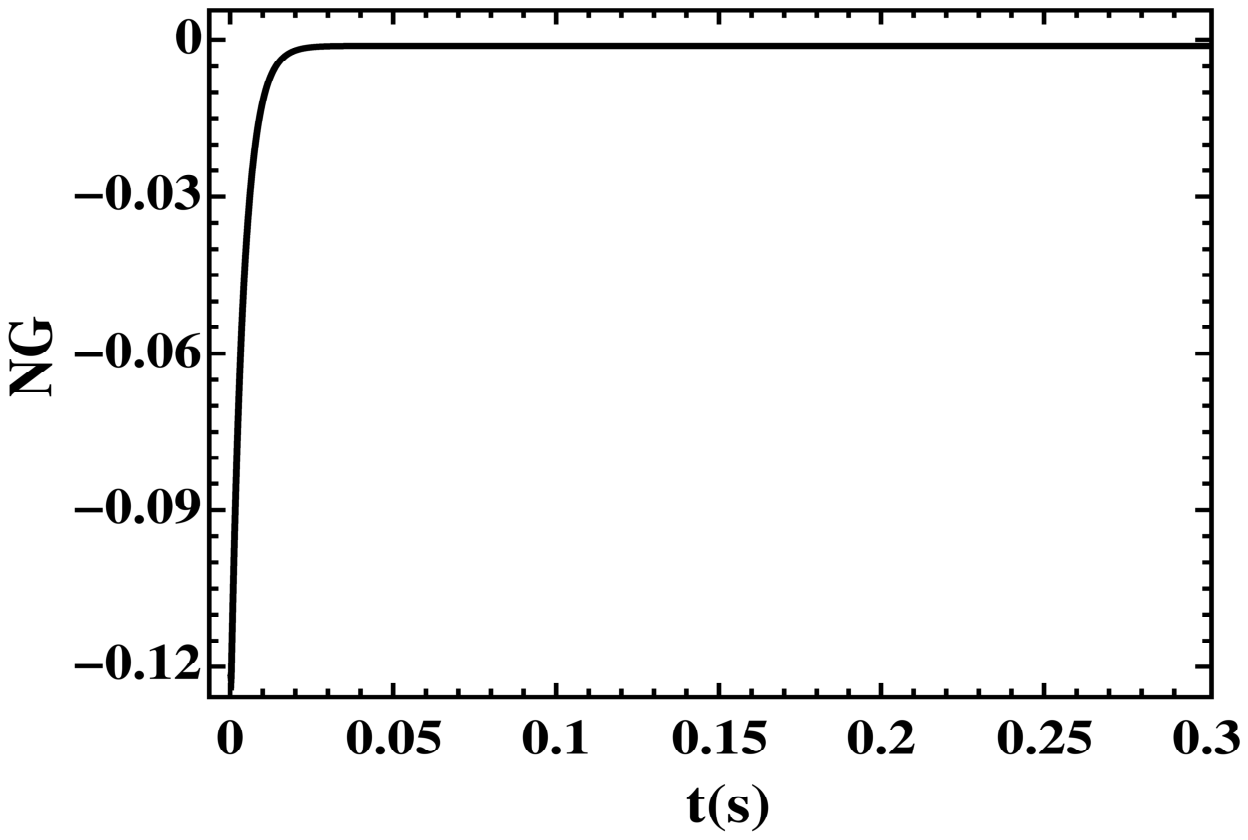}}
\caption{Plot of the non-Gaussianity $NG$ of Eq.~\protect\ref{eq:ng} versus time (soon after the cat state generation) for (a) $\bar{n}=100$ starting from the mechanical ground state, (b) $\bar{n}=10$ starting from the two-phonon cooling initial state; (c) $\bar{n}=100$ starting from the two-phonon cooling initial state. The other parameters are given in the text and coincide with those of Fig.~\protect\ref{fig:fid1}.}\label{fig:ngau}
\end{figure*}

We have calculated the quantity $NG$ quantifying non-Gaussianity by restricting to Gaussian unitary maps formed by a composition of the phase space displacement operator $D(\alpha)=\exp\left[\alpha b^{\dagger}-\alpha^* b\right]$ and of the squeezing operator $S(s)=\exp\left[(s/2)( b^{\dagger})^2-(s^*/2) b^2\right]$, and minimizing $NG$ over $\alpha$ and $s$. The values $\alpha = 0.35 i$ and $s =0.01$ work very well at all time instants after the cat state generation, either when starting from the mechanical ground state and when starting from the mixture of the vacuum and one phonon state obtained with two-phonon cooling. Plot of the time evolution of $NG$ soon after the cat state generation, in the three cases studied above, i.e., starting from the ground state and $\bar{n}=100$ (a), starting from two-phonon cooling and $\bar{n}=10$ (b), and $\bar{n}=100$ (c), are shown in Fig.~\ref{fig:ngau}. In all cases we see an exponential-like ``decay'' of non-Gaussianity to the Gaussian limit $NG = 0$, as expected, which is faster in the cases when $\bar{n}=100$; the non-Gaussianity decay rate is in good agreement with the usual decoherence rate $2\gamma_m |\beta|^2\left(2\bar{n}+1\right)$. Therefore the measure of non-Gaussianity of Eq.~(\ref{eq:ng}) proposed in Ref.~\cite{Palma2013} detects very well the non-Gaussian property, and for the present even cat state the dynamics of non-Gaussianity provides a satisfactory description of the decoherence process.

\section{Conclusions}

We have proposed a scheme for the deterministic generation of a linear superposition of two coherent states of a MR based on the implementation of an engineered reservoir realized by a bad cavity mode, bichromatically driven and coupled \emph{quadratically} with the MR. The proposal extends in various aspects the proposal of Ref.~\cite{Tan2013} and is feasible adopting either MIM optomechanical setups or levitated nanospheres trapped around an intensity maximum of the optical cavity mode. The interplay between the engineered reservoir and the natural thermal reservoir of the MR allows the efficient generation of the linear superposition state in a transient regime if the rate of the engineered reservoir $\Gamma$ is larger than $\gamma_m \bar{n}$, which is experimentally achievable in cryogenic environments at about $T \sim 10$ mK. The generation of an even superposition of two coherent states of opposite phases is almost ideal when starting from the MR ground state. This initial condition could be obtained by laser pre-cooling the MR through a linear optomechanical interaction, which however must be then suddenly switched to a quadratic interaction, by shifting for example the membrane to a node of the cavity mode. However the cat state generation is very efficient also when precooling is realized by exploiting only the two-phonon relaxation processes associated with the quadratic interaction~\cite{Nunnenkamp2010}, which is much easier to implement since it is based on the same configuration allowing the cat state generation.

At longer times, the thermal reservoir is responsible for the progressive decoherence of the generated superposition state, which asymptotically tends to a steady state given by the incoherent mixture of the two coherent states of the superposition, and which can be satisfactorily approximated by a simple analytical expression. For this reason, the present protocol is ideal for testing decoherence models acting on nanomechanical resonators.

An important issue is also the development of an efficient detection of the generated MR state. A satisfactory detection could be obtained by realizing a homodyne tomography~\cite{Dariano1994} of the Wigner function of the generated state. Homodyne tomography of the MR state could be obtained by first transferring such state to an auxiliary cavity mode, weakly linearly coupled to the MR, as suggested in Ref.~\cite{Vitali2007} or adopting the pulsed homodyne measurement scheme of Ref.~\cite{Vanner2011}. When the auxiliary cavity mode is driven on its first red sideband and can be adiabatically eliminated, its output field $a_2^{\rm out}$ is proportional to the MR annihilation operator $b$ plus additional noise~\cite{Vitali2007}, and therefore a calibrated homodyne detection of this output field at various phases could be exploited for a tomographic reconstruction of the MR Wigner function. The presence of the driven, weakly linearly coupled detection cavity mode affects the two-phonon processes creating the engineered reservoir, and therefore the detection process should be turned on only after the cat state generation has been completed.

\section{Acknowledgments}

This work has been supported by the European Commission (ITN-Marie Curie project cQOM), and by MIUR (PRIN 2011).

\bibliography{cat-state1}

\begin{thebibliography}{53}%
\makeatletter
\providecommand \@ifxundefined [1]{%
 \@ifx{#1\undefined}
}%
\providecommand \@ifnum [1]{%
 \ifnum #1\expandafter \@firstoftwo
 \else \expandafter \@secondoftwo
 \fi
}%
\providecommand \@ifx [1]{%
 \ifx #1\expandafter \@firstoftwo
 \else \expandafter \@secondoftwo
 \fi
}%
\providecommand \natexlab [1]{#1}%
\providecommand \enquote  [1]{``#1''}%
\providecommand \bibnamefont  [1]{#1}%
\providecommand \bibfnamefont [1]{#1}%
\providecommand \citenamefont [1]{#1}%
\providecommand \href@noop [0]{\@secondoftwo}%
\providecommand \href [0]{\begingroup \@sanitize@url \@href}%
\providecommand \@href[1]{\@@startlink{#1}\@@href}%
\providecommand \@@href[1]{\endgroup#1\@@endlink}%
\providecommand \@sanitize@url [0]{\catcode `\\12\catcode `\$12\catcode
  `\&12\catcode `\#12\catcode `\^12\catcode `\_12\catcode `\%12\relax}%
\providecommand \@@startlink[1]{}%
\providecommand \@@endlink[0]{}%
\providecommand \url  [0]{\begingroup\@sanitize@url \@url }%
\providecommand \@url [1]{\endgroup\@href {#1}{\urlprefix }}%
\providecommand \urlprefix  [0]{URL }%
\providecommand \Eprint [0]{\href }%
\@ifxundefined \urlstyle {%
  \providecommand \doi  [0]{\begingroup \@sanitize@url \@doi}%
  \providecommand \@doi [1]{\endgroup \@@startlink {\doibase
  #1}doi:\discretionary {}{}{}#1\@@endlink }%
}{%
  \providecommand \doi  [0]{doi:\discretionary{}{}{}\begingroup
  \urlstyle{rm}\Url }%
}%
\providecommand \doibase [0]{http://dx.doi.org/}%
\providecommand \Doi [0]{\begingroup \@sanitize@url \@Doi }%
\providecommand \@Doi  [1]{\endgroup\@@startlink{\doibase#1}\@@Doi}%
\providecommand \@@Doi [1]{#1\@@endlink}%
\providecommand \selectlanguage [0]{\@gobble}%
\providecommand \bibinfo  [0]{\@secondoftwo}%
\providecommand \bibfield  [0]{\@secondoftwo}%
\providecommand \translation [1]{[#1]}%
\providecommand \BibitemOpen [0]{}%
\providecommand \bibitemStop [0]{}%
\providecommand \bibitemNoStop [0]{.\EOS\space}%
\providecommand \EOS [0]{\spacefactor3000\relax}%
\providecommand \BibitemShut  [1]{\csname bibitem#1\endcsname}%
\bibitem [{\citenamefont {Diehl}\ \emph {et~al.}(2008)\citenamefont {Diehl},
  \citenamefont {Micheli}, \citenamefont {Kantian}, \citenamefont {Kraus},
  \citenamefont {Büchler},\ and\ \citenamefont {Zoller}}]{Diehl2008}%
  \BibitemOpen
  \bibfield  {author} {\bibinfo {author} {\bibfnamefont {S.}~\bibnamefont
  {Diehl}}, \bibinfo {author} {\bibfnamefont {A.}~\bibnamefont {Micheli}},
  \bibinfo {author} {\bibfnamefont {A.}~\bibnamefont {Kantian}}, \bibinfo
  {author} {\bibfnamefont {B.}~\bibnamefont {Kraus}}, \bibinfo {author}
  {\bibfnamefont {H.~P.}\ \bibnamefont {Büchler}}, \ and\ \bibinfo {author}
  {\bibfnamefont {P.}~\bibnamefont {Zoller}},\ }\href@noop {} {\bibfield
  {journal} {\bibinfo  {journal} {Nat. Phys.},\ }\textbf {\bibinfo {volume}
  {4}},\ \bibinfo {pages} {878} (\bibinfo {year} {2008})}\BibitemShut {NoStop}%
\bibitem [{\citenamefont {Verstraete}\ \emph {et~al.}(2009)\citenamefont
  {Verstraete}, \citenamefont {Wolf},\ and\ \citenamefont
  {Cirac}}]{Verstraete2009}%
  \BibitemOpen
  \bibfield  {author} {\bibinfo {author} {\bibfnamefont {F.}~\bibnamefont
  {Verstraete}}, \bibinfo {author} {\bibfnamefont {M.~M.}\ \bibnamefont
  {Wolf}}, \ and\ \bibinfo {author} {\bibfnamefont {J.~I.}\ \bibnamefont
  {Cirac}},\ }\href@noop {} {\bibfield  {journal} {\bibinfo  {journal} {Nat.
  Phys.},\ }\textbf {\bibinfo {volume} {5}},\ \bibinfo {pages} {633} (\bibinfo
  {year} {2009})}\BibitemShut {NoStop}%
\bibitem [{\citenamefont {Wineland}\ \emph {et~al.}(1978)\citenamefont
  {Wineland}, \citenamefont {Drullinger},\ and\ \citenamefont
  {Walls}}]{Wineland1978}%
  \BibitemOpen
  \bibfield  {author} {\bibinfo {author} {\bibfnamefont {D.}~\bibnamefont
  {Wineland}}, \bibinfo {author} {\bibfnamefont {R.}~\bibnamefont
  {Drullinger}}, \ and\ \bibinfo {author} {\bibfnamefont {F.}~\bibnamefont
  {Walls}},\ }\href@noop {} {\bibfield  {journal} {\bibinfo  {journal} {Phys.
  Rev. Lett.},\ }\textbf {\bibinfo {volume} {40}},\ \bibinfo {pages} {1639}
  (\bibinfo {year} {1978})}\BibitemShut {NoStop}%
\bibitem [{\citenamefont {Poyatos}\ \emph {et~al.}(1996)\citenamefont
  {Poyatos}, \citenamefont {Cirac},\ and\ \citenamefont
  {Zoller}}]{Poyatos1996}%
  \BibitemOpen
  \bibfield  {author} {\bibinfo {author} {\bibfnamefont {J.}~\bibnamefont
  {Poyatos}}, \bibinfo {author} {\bibfnamefont {J.}~\bibnamefont {Cirac}}, \
  and\ \bibinfo {author} {\bibfnamefont {P.}~\bibnamefont {Zoller}},\
  }\href@noop {} {\bibfield  {journal} {\bibinfo  {journal} {Phys. Rev.
  Lett.},\ }\textbf {\bibinfo {volume} {77}},\ \bibinfo {pages} {4728}
  (\bibinfo {year} {1996})}\BibitemShut {NoStop}%
\bibitem [{\citenamefont {Carvalho}\ \emph {et~al.}(2001)\citenamefont
  {Carvalho}, \citenamefont {Milman}, \citenamefont {de~Matos~Filho},\ and\
  \citenamefont {Davidovich}}]{Carvalho2001}%
  \BibitemOpen
  \bibfield  {author} {\bibinfo {author} {\bibfnamefont {A.~R.~R.}\
  \bibnamefont {Carvalho}}, \bibinfo {author} {\bibfnamefont {P.}~\bibnamefont
  {Milman}}, \bibinfo {author} {\bibfnamefont {R.~L.}\ \bibnamefont
  {de~Matos~Filho}}, \ and\ \bibinfo {author} {\bibfnamefont {L.}~\bibnamefont
  {Davidovich}},\ }\href@noop {} {\bibfield  {journal} {\bibinfo  {journal}
  {Phys. Rev. Lett.},\ }\textbf {\bibinfo {volume} {86}},\ \bibinfo {pages}
  {4988} (\bibinfo {year} {2001})}\BibitemShut {NoStop}%
\bibitem [{\citenamefont {Syassen}\ \emph {et~al.}(2008)\citenamefont
  {Syassen}, \citenamefont {Bauer}, \citenamefont {Lettner}, \citenamefont
  {Volz}, \citenamefont {Dietze}, \citenamefont {Garcia-Ripoll}, \citenamefont
  {Cirac}, \citenamefont {Rempe},\ and\ \citenamefont {D\"urr}}]{Syassen2008}%
  \BibitemOpen
  \bibfield  {author} {\bibinfo {author} {\bibfnamefont {N.}~\bibnamefont
  {Syassen}}, \bibinfo {author} {\bibfnamefont {D.}~\bibnamefont {Bauer}},
  \bibinfo {author} {\bibfnamefont {M.}~\bibnamefont {Lettner}}, \bibinfo
  {author} {\bibfnamefont {T.}~\bibnamefont {Volz}}, \bibinfo {author}
  {\bibfnamefont {D.}~\bibnamefont {Dietze}}, \bibinfo {author} {\bibfnamefont
  {J.}~\bibnamefont {Garcia-Ripoll}}, \bibinfo {author} {\bibfnamefont
  {J.}~\bibnamefont {Cirac}}, \bibinfo {author} {\bibfnamefont
  {G.}~\bibnamefont {Rempe}}, \ and\ \bibinfo {author} {\bibfnamefont
  {S.}~\bibnamefont {D\"urr}},\ }\href@noop {} {\bibfield  {journal} {\bibinfo
  {journal} {Science},\ }\textbf {\bibinfo {volume} {320}},\ \bibinfo {pages}
  {1329} (\bibinfo {year} {2008})}\BibitemShut {NoStop}%
\bibitem [{\citenamefont {Krauter}\ \emph {et~al.}(2011)\citenamefont
  {Krauter}, \citenamefont {Muschik}, \citenamefont {Jensen}, \citenamefont
  {Wasilewski}, \citenamefont {Petersen}, \citenamefont {Cirac},\ and\
  \citenamefont {Polzik}}]{Krauter2011}%
  \BibitemOpen
  \bibfield  {author} {\bibinfo {author} {\bibfnamefont {H.}~\bibnamefont
  {Krauter}}, \bibinfo {author} {\bibfnamefont {C.~A.}\ \bibnamefont
  {Muschik}}, \bibinfo {author} {\bibfnamefont {K.}~\bibnamefont {Jensen}},
  \bibinfo {author} {\bibfnamefont {W.}~\bibnamefont {Wasilewski}}, \bibinfo
  {author} {\bibfnamefont {J.~M.}\ \bibnamefont {Petersen}}, \bibinfo {author}
  {\bibfnamefont {J.~I.}\ \bibnamefont {Cirac}}, \ and\ \bibinfo {author}
  {\bibfnamefont {E.~S.}\ \bibnamefont {Polzik}},\ }\href@noop {} {\bibfield
  {journal} {\bibinfo  {journal} {Phys. Rev. Lett.},\ }\textbf {\bibinfo
  {volume} {107}},\ \bibinfo {pages} {080503} (\bibinfo {year}
  {2011})}\BibitemShut {NoStop}%
\bibitem [{\citenamefont {Pielawa}\ \emph {et~al.}(2007)\citenamefont
  {Pielawa}, \citenamefont {Morigi}, \citenamefont {Vitali},\ and\
  \citenamefont {Davidovich}}]{Pielawa2007}%
  \BibitemOpen
  \bibfield  {author} {\bibinfo {author} {\bibfnamefont {S.}~\bibnamefont
  {Pielawa}}, \bibinfo {author} {\bibfnamefont {G.}~\bibnamefont {Morigi}},
  \bibinfo {author} {\bibfnamefont {D.}~\bibnamefont {Vitali}}, \ and\ \bibinfo
  {author} {\bibfnamefont {L.}~\bibnamefont {Davidovich}},\ }\href@noop {}
  {\bibfield  {journal} {\bibinfo  {journal} {Phys. Rev. Lett.},\ }\textbf
  {\bibinfo {volume} {98}},\ \bibinfo {pages} {240401} (\bibinfo {year}
  {2007})}\BibitemShut {NoStop}%
\bibitem [{\citenamefont {Pielawa}\ \emph {et~al.}(2010)\citenamefont
  {Pielawa}, \citenamefont {Davidovich}, \citenamefont {Vitali},\ and\
  \citenamefont {Morigi}}]{Pielawa2010}%
  \BibitemOpen
  \bibfield  {author} {\bibinfo {author} {\bibfnamefont {S.}~\bibnamefont
  {Pielawa}}, \bibinfo {author} {\bibfnamefont {L.}~\bibnamefont {Davidovich}},
  \bibinfo {author} {\bibfnamefont {D.}~\bibnamefont {Vitali}}, \ and\ \bibinfo
  {author} {\bibfnamefont {G.}~\bibnamefont {Morigi}},\ }\href@noop {}
  {\bibfield  {journal} {\bibinfo  {journal} {Phys. Rev. A},\ }\textbf
  {\bibinfo {volume} {81}},\ \bibinfo {pages} {043802} (\bibinfo {year}
  {2010})}\BibitemShut {NoStop}%
\bibitem [{\citenamefont {Wang}\ and\ \citenamefont {Clerk}(2013)}]{Wang2013}%
  \BibitemOpen
  \bibfield  {author} {\bibinfo {author} {\bibfnamefont {Y.-D.}\ \bibnamefont
  {Wang}}\ and\ \bibinfo {author} {\bibfnamefont {A.~A.}\ \bibnamefont
  {Clerk}},\ }\href@noop {} {\bibfield  {journal} {\bibinfo  {journal} {Phys.
  Rev. Lett.},\ }\textbf {\bibinfo {volume} {110}},\ \bibinfo {pages} {253601}
  (\bibinfo {year} {2013})}\BibitemShut {NoStop}%
\bibitem [{\citenamefont {Tan}\ \emph {et~al.}(2013){\natexlab{a}}\citenamefont
  {Tan}, \citenamefont {Li},\ and\ \citenamefont {Meystre}}]{Tan2013a}%
  \BibitemOpen
  \bibfield  {author} {\bibinfo {author} {\bibfnamefont {H.}~\bibnamefont
  {Tan}}, \bibinfo {author} {\bibfnamefont {G.}~\bibnamefont {Li}}, \ and\
  \bibinfo {author} {\bibfnamefont {P.}~\bibnamefont {Meystre}},\ }\href@noop
  {} {\bibfield  {journal} {\bibinfo  {journal} {Phys. Rev. A},\ }\textbf
  {\bibinfo {volume} {87}},\ \bibinfo {pages} {033829} (\bibinfo {year}
  {2013}{\natexlab{a}})}\BibitemShut {NoStop}%
\bibitem [{\citenamefont {Bose}\ \emph {et~al.}(1999)\citenamefont {Bose},
  \citenamefont {Jacobs},\ and\ \citenamefont {Knight}}]{Bose1999}%
  \BibitemOpen
  \bibfield  {author} {\bibinfo {author} {\bibfnamefont {S.}~\bibnamefont
  {Bose}}, \bibinfo {author} {\bibfnamefont {K.}~\bibnamefont {Jacobs}}, \ and\
  \bibinfo {author} {\bibfnamefont {P.~L.}\ \bibnamefont {Knight}},\
  }\href@noop {} {\bibfield  {journal} {\bibinfo  {journal} {Phys. Rev. A},\
  }\textbf {\bibinfo {volume} {59}},\ \bibinfo {pages} {3204} (\bibinfo {year}
  {1999})}\BibitemShut {NoStop}%
\bibitem [{\citenamefont {Marshall}\ \emph {et~al.}(2003)\citenamefont
  {Marshall}, \citenamefont {Simon}, \citenamefont {Penrose},\ and\
  \citenamefont {Bouwmeester}}]{Marshall2003}%
  \BibitemOpen
  \bibfield  {author} {\bibinfo {author} {\bibfnamefont {W.}~\bibnamefont
  {Marshall}}, \bibinfo {author} {\bibfnamefont {C.}~\bibnamefont {Simon}},
  \bibinfo {author} {\bibfnamefont {R.}~\bibnamefont {Penrose}}, \ and\
  \bibinfo {author} {\bibfnamefont {D.}~\bibnamefont {Bouwmeester}},\
  }\href@noop {} {\bibfield  {journal} {\bibinfo  {journal} {Phys. Rev.
  Lett.},\ }\textbf {\bibinfo {volume} {91}},\ \bibinfo {pages} {130401}
  (\bibinfo {year} {2003})}\BibitemShut {NoStop}%
\bibitem [{\citenamefont {Paternostro}(2011)}]{Paternostro2011}%
  \BibitemOpen
  \bibfield  {author} {\bibinfo {author} {\bibfnamefont {M.}~\bibnamefont
  {Paternostro}},\ }\href@noop {} {\bibfield  {journal} {\bibinfo  {journal}
  {Phys. Rev. Lett.},\ }\textbf {\bibinfo {volume} {106}},\ \bibinfo {pages}
  {183601} (\bibinfo {year} {2011})}\BibitemShut {NoStop}%
\bibitem [{\citenamefont {Vanner}\ \emph {et~al.}(2013)\citenamefont {Vanner},
  \citenamefont {Aspelmeyer},\ and\ \citenamefont {Kim}}]{Vanner2013}%
  \BibitemOpen
  \bibfield  {author} {\bibinfo {author} {\bibfnamefont {M.~R.}\ \bibnamefont
  {Vanner}}, \bibinfo {author} {\bibfnamefont {M.}~\bibnamefont {Aspelmeyer}},
  \ and\ \bibinfo {author} {\bibfnamefont {M.~S.}\ \bibnamefont {Kim}},\
  }\href@noop {} {\bibfield  {journal} {\bibinfo  {journal} {Phys. Rev.
  Lett.},\ }\textbf {\bibinfo {volume} {110}},\ \bibinfo {pages} {010504}
  (\bibinfo {year} {2013})}\BibitemShut {NoStop}%
\bibitem [{\citenamefont {Romero-Isart}\ \emph
  {et~al.}(2011){\natexlab{a}}\citenamefont {Romero-Isart}, \citenamefont
  {Pflanzer}, \citenamefont {Blaser}, \citenamefont {Kaltenbaek}, \citenamefont
  {Kiesel}, \citenamefont {Aspelmeyer},\ and\ \citenamefont
  {Cirac}}]{Romero-Isart2011a}%
  \BibitemOpen
  \bibfield  {author} {\bibinfo {author} {\bibfnamefont {O.}~\bibnamefont
  {Romero-Isart}}, \bibinfo {author} {\bibfnamefont {A.~C.}\ \bibnamefont
  {Pflanzer}}, \bibinfo {author} {\bibfnamefont {F.}~\bibnamefont {Blaser}},
  \bibinfo {author} {\bibfnamefont {R.}~\bibnamefont {Kaltenbaek}}, \bibinfo
  {author} {\bibfnamefont {N.}~\bibnamefont {Kiesel}}, \bibinfo {author}
  {\bibfnamefont {M.}~\bibnamefont {Aspelmeyer}}, \ and\ \bibinfo {author}
  {\bibfnamefont {J.~I.}\ \bibnamefont {Cirac}},\ }\href@noop {} {\bibfield
  {journal} {\bibinfo  {journal} {Phys. Rev. Lett.},\ }\textbf {\bibinfo
  {volume} {107}},\ \bibinfo {pages} {020405} (\bibinfo {year}
  {2011}{\natexlab{a}})}\BibitemShut {NoStop}%
\bibitem [{\citenamefont {Romero-Isart}\ \emph
  {et~al.}(2011){\natexlab{b}}\citenamefont {Romero-Isart}, \citenamefont
  {Pflanzer}, \citenamefont {Juan}, \citenamefont {Quidant}, \citenamefont
  {Kiesel}, \citenamefont {Aspelmeyer},\ and\ \citenamefont
  {Cirac}}]{Romero-Isart2011}%
  \BibitemOpen
  \bibfield  {author} {\bibinfo {author} {\bibfnamefont {O.}~\bibnamefont
  {Romero-Isart}}, \bibinfo {author} {\bibfnamefont {A.~C.}\ \bibnamefont
  {Pflanzer}}, \bibinfo {author} {\bibfnamefont {M.~L.}\ \bibnamefont {Juan}},
  \bibinfo {author} {\bibfnamefont {R.}~\bibnamefont {Quidant}}, \bibinfo
  {author} {\bibfnamefont {N.}~\bibnamefont {Kiesel}}, \bibinfo {author}
  {\bibfnamefont {M.}~\bibnamefont {Aspelmeyer}}, \ and\ \bibinfo {author}
  {\bibfnamefont {J.~I.}\ \bibnamefont {Cirac}},\ }\href@noop {} {\bibfield
  {journal} {\bibinfo  {journal} {Phys. Rev. A},\ }\textbf {\bibinfo {volume}
  {83}},\ \bibinfo {pages} {013803} (\bibinfo {year}
  {2011}{\natexlab{b}})}\BibitemShut {NoStop}%
\bibitem [{\citenamefont {Jacobs}\ \emph {et~al.}(2009)\citenamefont {Jacobs},
  \citenamefont {Tian},\ and\ \citenamefont {Finn}}]{Jacobs2009}%
  \BibitemOpen
  \bibfield  {author} {\bibinfo {author} {\bibfnamefont {K.}~\bibnamefont
  {Jacobs}}, \bibinfo {author} {\bibfnamefont {L.}~\bibnamefont {Tian}}, \ and\
  \bibinfo {author} {\bibfnamefont {J.}~\bibnamefont {Finn}},\ }\href@noop {}
  {\bibfield  {journal} {\bibinfo  {journal} {Phys. Rev. Lett.},\ }\textbf
  {\bibinfo {volume} {102}},\ \bibinfo {pages} {057208} (\bibinfo {year}
  {2009})}\BibitemShut {NoStop}%
\bibitem [{\citenamefont {Jacobs}\ \emph {et~al.}(2011)\citenamefont {Jacobs},
  \citenamefont {Finn},\ and\ \citenamefont {Vinjanampathy}}]{Jacobs2011}%
  \BibitemOpen
  \bibfield  {author} {\bibinfo {author} {\bibfnamefont {K.}~\bibnamefont
  {Jacobs}}, \bibinfo {author} {\bibfnamefont {J.}~\bibnamefont {Finn}}, \ and\
  \bibinfo {author} {\bibfnamefont {S.}~\bibnamefont {Vinjanampathy}},\
  }\href@noop {} {\bibfield  {journal} {\bibinfo  {journal} {Phys. rev.
  Lett.},\ }\textbf {\bibinfo {volume} {83}},\ \bibinfo {pages} {041801(R)}
  (\bibinfo {year} {2011})}\BibitemShut {NoStop}%
\bibitem [{\citenamefont {Tan}\ \emph {et~al.}(2013){\natexlab{b}}\citenamefont
  {Tan}, \citenamefont {Bariani}, \citenamefont {Li},\ and\ \citenamefont
  {Meystre}}]{Tan2013}%
  \BibitemOpen
  \bibfield  {author} {\bibinfo {author} {\bibfnamefont {H.}~\bibnamefont
  {Tan}}, \bibinfo {author} {\bibfnamefont {F.}~\bibnamefont {Bariani}},
  \bibinfo {author} {\bibfnamefont {G.}~\bibnamefont {Li}}, \ and\ \bibinfo
  {author} {\bibfnamefont {P.}~\bibnamefont {Meystre}},\ }\href@noop {}
  {\bibfield  {journal} {\bibinfo  {journal} {Phys. Rev. A},\ }\textbf
  {\bibinfo {volume} {88}},\ \bibinfo {pages} {023817} (\bibinfo {year}
  {2013}{\natexlab{b}})}\BibitemShut {NoStop}%
\bibitem [{\citenamefont {Del\'eglise}\ \emph {et~al.}(2008)\citenamefont
  {Del\'eglise}, \citenamefont {Dotsenko}, \citenamefont {Sayrin},
  \citenamefont {Bernu}, \citenamefont {Brune}, \citenamefont {Raimond},\ and\
  \citenamefont {Haroche}}]{Deleglise2008}%
  \BibitemOpen
  \bibfield  {author} {\bibinfo {author} {\bibfnamefont {S.}~\bibnamefont
  {Del\'eglise}}, \bibinfo {author} {\bibfnamefont {I.}~\bibnamefont
  {Dotsenko}}, \bibinfo {author} {\bibfnamefont {C.}~\bibnamefont {Sayrin}},
  \bibinfo {author} {\bibfnamefont {J.}~\bibnamefont {Bernu}}, \bibinfo
  {author} {\bibfnamefont {M.}~\bibnamefont {Brune}}, \bibinfo {author}
  {\bibfnamefont {J.-M.}\ \bibnamefont {Raimond}}, \ and\ \bibinfo {author}
  {\bibfnamefont {S.}~\bibnamefont {Haroche}},\ }\href@noop {} {\bibfield
  {journal} {\bibinfo  {journal} {Nature (London)},\ }\textbf {\bibinfo
  {volume} {455}},\ \bibinfo {pages} {510} (\bibinfo {year}
  {2008})}\BibitemShut {NoStop}%
\bibitem [{\citenamefont {Myatt}\ \emph {et~al.}(2000)\citenamefont {Myatt},
  \citenamefont {King}, \citenamefont {Turchette}, \citenamefont {Sackett},
  \citenamefont {Kielpinski}, \citenamefont {Itano}, \citenamefont {Monroe},\
  and\ \citenamefont {Wineland}}]{Myatt2000}%
  \BibitemOpen
  \bibfield  {author} {\bibinfo {author} {\bibfnamefont {C.~J.}\ \bibnamefont
  {Myatt}}, \bibinfo {author} {\bibfnamefont {B.~E.}\ \bibnamefont {King}},
  \bibinfo {author} {\bibfnamefont {Q.~A.}\ \bibnamefont {Turchette}}, \bibinfo
  {author} {\bibfnamefont {C.~A.}\ \bibnamefont {Sackett}}, \bibinfo {author}
  {\bibfnamefont {D.}~\bibnamefont {Kielpinski}}, \bibinfo {author}
  {\bibfnamefont {W.~M.}\ \bibnamefont {Itano}}, \bibinfo {author}
  {\bibfnamefont {C.}~\bibnamefont {Monroe}}, \ and\ \bibinfo {author}
  {\bibfnamefont {D.~J.}\ \bibnamefont {Wineland}},\ }\href@noop {} {\bibfield
  {journal} {\bibinfo  {journal} {Nature (London)},\ }\textbf {\bibinfo
  {volume} {403}},\ \bibinfo {pages} {269} (\bibinfo {year}
  {2000})}\BibitemShut {NoStop}%
\bibitem [{\citenamefont {Thompson}\ \emph {et~al.}(2008)\citenamefont
  {Thompson}, \citenamefont {Zwickl}, \citenamefont {Jayich}, \citenamefont
  {Marquardt}, \citenamefont {Girvin},\ and\ \citenamefont
  {Harris}}]{Thompson2008}%
  \BibitemOpen
  \bibfield  {author} {\bibinfo {author} {\bibfnamefont {J.~D.}\ \bibnamefont
  {Thompson}}, \bibinfo {author} {\bibfnamefont {B.~M.}\ \bibnamefont
  {Zwickl}}, \bibinfo {author} {\bibfnamefont {A.~M.}\ \bibnamefont {Jayich}},
  \bibinfo {author} {\bibfnamefont {F.}~\bibnamefont {Marquardt}}, \bibinfo
  {author} {\bibfnamefont {S.~M.}\ \bibnamefont {Girvin}}, \ and\ \bibinfo
  {author} {\bibfnamefont {J.~G.~E.}\ \bibnamefont {Harris}},\ }\href@noop {}
  {\bibfield  {journal} {\bibinfo  {journal} {Nature (London)},\ }\textbf
  {\bibinfo {volume} {452}},\ \bibinfo {pages} {72} (\bibinfo {year}
  {2008})}\BibitemShut {NoStop}%
\bibitem [{\citenamefont {Sankey}\ \emph {et~al.}(2010)\citenamefont {Sankey},
  \citenamefont {Yang}, \citenamefont {Zwickl}, \citenamefont {Jayich},\ and\
  \citenamefont {Harris}}]{Sankey2010}%
  \BibitemOpen
  \bibfield  {author} {\bibinfo {author} {\bibfnamefont {J.~C.}\ \bibnamefont
  {Sankey}}, \bibinfo {author} {\bibfnamefont {C.}~\bibnamefont {Yang}},
  \bibinfo {author} {\bibfnamefont {B.~M.}\ \bibnamefont {Zwickl}}, \bibinfo
  {author} {\bibfnamefont {A.~M.}\ \bibnamefont {Jayich}}, \ and\ \bibinfo
  {author} {\bibfnamefont {J.~G.~E.}\ \bibnamefont {Harris}},\ }\href@noop {}
  {\bibfield  {journal} {\bibinfo  {journal} {Nat. Phys.},\ }\textbf {\bibinfo
  {volume} {6}},\ \bibinfo {pages} {707} (\bibinfo {year} {2010})}\BibitemShut
  {NoStop}%
\bibitem [{\citenamefont {Karuza}\ \emph {et~al.}(2013)\citenamefont {Karuza},
  \citenamefont {Galassi}, \citenamefont {Biancofiore}, \citenamefont
  {Molinelli}, \citenamefont {Natali}, \citenamefont {Tombesi}, \citenamefont
  {{Di Giuseppe}},\ and\ \citenamefont {Vitali}}]{Karuza2013}%
  \BibitemOpen
  \bibfield  {author} {\bibinfo {author} {\bibfnamefont {M.}~\bibnamefont
  {Karuza}}, \bibinfo {author} {\bibfnamefont {M.}~\bibnamefont {Galassi}},
  \bibinfo {author} {\bibfnamefont {C.}~\bibnamefont {Biancofiore}}, \bibinfo
  {author} {\bibfnamefont {C.}~\bibnamefont {Molinelli}}, \bibinfo {author}
  {\bibfnamefont {R.}~\bibnamefont {Natali}}, \bibinfo {author} {\bibfnamefont
  {P.}~\bibnamefont {Tombesi}}, \bibinfo {author} {\bibfnamefont
  {G.}~\bibnamefont {{Di Giuseppe}}}, \ and\ \bibinfo {author} {\bibfnamefont
  {D.}~\bibnamefont {Vitali}},\ }\href@noop {} {\bibfield  {journal} {\bibinfo
  {journal} {J. Opt.},\ }\textbf {\bibinfo {volume} {15}},\ \bibinfo {pages}
  {025704} (\bibinfo {year} {2013})}\BibitemShut {NoStop}%
\bibitem [{\citenamefont {Barker}(2010)}]{Barker2010}%
  \BibitemOpen
  \bibfield  {author} {\bibinfo {author} {\bibfnamefont {P.~F.}\ \bibnamefont
  {Barker}},\ }\href@noop {} {\bibfield  {journal} {\bibinfo  {journal} {Phys.
  Rev. Lett.},\ }\textbf {\bibinfo {volume} {105}},\ \bibinfo {pages} {073002}
  (\bibinfo {year} {2010})}\BibitemShut {NoStop}%
\bibitem [{\citenamefont {Li}\ \emph {et~al.}(2011)\citenamefont {Li},
  \citenamefont {Kheifets},\ and\ \citenamefont {Raizen}}]{Li2011}%
  \BibitemOpen
  \bibfield  {author} {\bibinfo {author} {\bibfnamefont {T.}~\bibnamefont
  {Li}}, \bibinfo {author} {\bibfnamefont {S.}~\bibnamefont {Kheifets}}, \ and\
  \bibinfo {author} {\bibfnamefont {M.~G.}\ \bibnamefont {Raizen}},\
  }\href@noop {} {\bibfield  {journal} {\bibinfo  {journal} {Nat. Phys.},\
  }\textbf {\bibinfo {volume} {7}},\ \bibinfo {pages} {527} (\bibinfo {year}
  {2011})}\BibitemShut {NoStop}%
\bibitem [{\citenamefont {Gieseler}\ \emph {et~al.}(2012)\citenamefont
  {Gieseler}, \citenamefont {Deutsch}, \citenamefont {Quidant},\ and\
  \citenamefont {Novotny}}]{Gieseler2012}%
  \BibitemOpen
  \bibfield  {author} {\bibinfo {author} {\bibfnamefont {J.}~\bibnamefont
  {Gieseler}}, \bibinfo {author} {\bibfnamefont {B.}~\bibnamefont {Deutsch}},
  \bibinfo {author} {\bibfnamefont {R.}~\bibnamefont {Quidant}}, \ and\
  \bibinfo {author} {\bibfnamefont {L.}~\bibnamefont {Novotny}},\ }\href@noop
  {} {\bibfield  {journal} {\bibinfo  {journal} {Phys. Rev. Lett.},\ }\textbf
  {\bibinfo {volume} {109}},\ \bibinfo {pages} {103603} (\bibinfo {year}
  {2012})}\BibitemShut {NoStop}%
\bibitem [{\citenamefont {Kiesel}\ \emph {et~al.}(2013)\citenamefont {Kiesel},
  \citenamefont {Blaser}, \citenamefont {Deli\'c}, \citenamefont {Grass},
  \citenamefont {Kaltenbaek},\ and\ \citenamefont {Aspelmeyer}}]{Kiesel2013}%
  \BibitemOpen
  \bibfield  {author} {\bibinfo {author} {\bibfnamefont {N.}~\bibnamefont
  {Kiesel}}, \bibinfo {author} {\bibfnamefont {F.}~\bibnamefont {Blaser}},
  \bibinfo {author} {\bibfnamefont {U.}~\bibnamefont {Deli\'c}}, \bibinfo
  {author} {\bibfnamefont {D.}~\bibnamefont {Grass}}, \bibinfo {author}
  {\bibfnamefont {R.}~\bibnamefont {Kaltenbaek}}, \ and\ \bibinfo {author}
  {\bibfnamefont {M.}~\bibnamefont {Aspelmeyer}},\ }\href@noop {} {\bibfield
  {journal} {\bibinfo  {journal} {arXiv:1304.6679v1 [quant-ph]}} (\bibinfo
  {year} {2013})}\BibitemShut {NoStop}%
\bibitem [{\citenamefont {Gardiner}\ and\ \citenamefont
  {Zoller}(2000)}]{Gardiner2000}%
  \BibitemOpen
  \bibfield  {author} {\bibinfo {author} {\bibfnamefont {C.~W.}\ \bibnamefont
  {Gardiner}}\ and\ \bibinfo {author} {\bibfnamefont {P.}~\bibnamefont
  {Zoller}},\ }\href@noop {} {\emph {\bibinfo {title} {Quantum Noise}}}\
  (\bibinfo  {publisher} {Springer},\ \bibinfo {year} {2000})\BibitemShut
  {NoStop}%
\bibitem [{\citenamefont {Wiseman}\ and\ \citenamefont
  {Milburn}(1993)}]{Wiseman1993}%
  \BibitemOpen
  \bibfield  {author} {\bibinfo {author} {\bibfnamefont {H.}~\bibnamefont
  {Wiseman}}\ and\ \bibinfo {author} {\bibfnamefont {G.~J.}\ \bibnamefont
  {Milburn}},\ }\href@noop {} {\bibfield  {journal} {\bibinfo  {journal} {Phys.
  Rev. A},\ }\textbf {\bibinfo {volume} {47}},\ \bibinfo {pages} {642}
  (\bibinfo {year} {1993})}\BibitemShut {NoStop}%
\bibitem [{\citenamefont {Wilson}\ \emph {et~al.}(2009)\citenamefont {Wilson},
  \citenamefont {Regal}, \citenamefont {Papp},\ and\ \citenamefont
  {Kimble}}]{Wilson2009}%
  \BibitemOpen
  \bibfield  {author} {\bibinfo {author} {\bibfnamefont {D.~J.}\ \bibnamefont
  {Wilson}}, \bibinfo {author} {\bibfnamefont {C.~A.}\ \bibnamefont {Regal}},
  \bibinfo {author} {\bibfnamefont {S.~B.}\ \bibnamefont {Papp}}, \ and\
  \bibinfo {author} {\bibfnamefont {H.~J.}\ \bibnamefont {Kimble}},\
  }\href@noop {} {\bibfield  {journal} {\bibinfo  {journal} {Phys. Rev.
  Lett.},\ }\textbf {\bibinfo {volume} {103}},\ \bibinfo {pages} {207204}
  (\bibinfo {year} {2009})}\BibitemShut {NoStop}%
\bibitem [{\citenamefont {Purdy}\ \emph {et~al.}(2012)\citenamefont {Purdy},
  \citenamefont {Peterson}, \citenamefont {Yu},\ and\ \citenamefont
  {Regal}}]{Purdy2012}%
  \BibitemOpen
  \bibfield  {author} {\bibinfo {author} {\bibfnamefont {T.~P.}\ \bibnamefont
  {Purdy}}, \bibinfo {author} {\bibfnamefont {R.~W.}\ \bibnamefont {Peterson}},
  \bibinfo {author} {\bibfnamefont {P.-L.}\ \bibnamefont {Yu}}, \ and\ \bibinfo
  {author} {\bibfnamefont {C.~A.}\ \bibnamefont {Regal}},\ }\href@noop {}
  {\bibfield  {journal} {\bibinfo  {journal} {New J. Phys.},\ }\textbf
  {\bibinfo {volume} {14}},\ \bibinfo {pages} {115021} (\bibinfo {year}
  {2012})}\BibitemShut {NoStop}%
\bibitem [{\citenamefont {Flowers-Jacobs}\ \emph {et~al.}(2012)\citenamefont
  {Flowers-Jacobs}, \citenamefont {Hoch}, \citenamefont {Sankey}, \citenamefont
  {Kashkanova}, \citenamefont {Jayich}, \citenamefont {Deutsch}, \citenamefont
  {Reichel},\ and\ \citenamefont {Harris}}]{Flowers-Jacobs2012}%
  \BibitemOpen
  \bibfield  {author} {\bibinfo {author} {\bibfnamefont {N.~E.}\ \bibnamefont
  {Flowers-Jacobs}}, \bibinfo {author} {\bibfnamefont {S.~W.}\ \bibnamefont
  {Hoch}}, \bibinfo {author} {\bibfnamefont {J.~C.}\ \bibnamefont {Sankey}},
  \bibinfo {author} {\bibfnamefont {A.}~\bibnamefont {Kashkanova}}, \bibinfo
  {author} {\bibfnamefont {A.~M.}\ \bibnamefont {Jayich}}, \bibinfo {author}
  {\bibfnamefont {C.}~\bibnamefont {Deutsch}}, \bibinfo {author} {\bibfnamefont
  {J.}~\bibnamefont {Reichel}}, \ and\ \bibinfo {author} {\bibfnamefont
  {J.~G.~E.}\ \bibnamefont {Harris}},\ }\href@noop {} {\bibfield  {journal}
  {\bibinfo  {journal} {Appl. Phys. Lett.},\ }\textbf {\bibinfo {volume}
  {101}},\ \bibinfo {pages} {221109} (\bibinfo {year} {2012})}\BibitemShut
  {NoStop}%
\bibitem [{\citenamefont {Karuza}\ \emph {et~al.}(2012)\citenamefont {Karuza},
  \citenamefont {Molinelli}, \citenamefont {Galassi}, \citenamefont
  {Biancofiore}, \citenamefont {Natali}, \citenamefont {Tombesi}, \citenamefont
  {{Di Giuseppe}},\ and\ \citenamefont {Vitali}}]{Karuza2012}%
  \BibitemOpen
  \bibfield  {author} {\bibinfo {author} {\bibfnamefont {M.}~\bibnamefont
  {Karuza}}, \bibinfo {author} {\bibfnamefont {C.}~\bibnamefont {Molinelli}},
  \bibinfo {author} {\bibfnamefont {M.}~\bibnamefont {Galassi}}, \bibinfo
  {author} {\bibfnamefont {C.}~\bibnamefont {Biancofiore}}, \bibinfo {author}
  {\bibfnamefont {R.}~\bibnamefont {Natali}}, \bibinfo {author} {\bibfnamefont
  {P.}~\bibnamefont {Tombesi}}, \bibinfo {author} {\bibfnamefont
  {G.}~\bibnamefont {{Di Giuseppe}}}, \ and\ \bibinfo {author} {\bibfnamefont
  {D.}~\bibnamefont {Vitali}},\ }\href@noop {} {\bibfield  {journal} {\bibinfo
  {journal} {New J. Phys.},\ }\textbf {\bibinfo {volume} {14}},\ \bibinfo
  {pages} {095015} (\bibinfo {year} {2012})}\BibitemShut {NoStop}%
\bibitem [{\citenamefont {Purdy}\ \emph {et~al.}(2013)\citenamefont {Purdy},
  \citenamefont {Peterson},\ and\ \citenamefont {Regal}}]{Purdy2013}%
  \BibitemOpen
  \bibfield  {author} {\bibinfo {author} {\bibfnamefont {T.~P.}\ \bibnamefont
  {Purdy}}, \bibinfo {author} {\bibfnamefont {R.~W.}\ \bibnamefont {Peterson}},
  \ and\ \bibinfo {author} {\bibfnamefont {C.~A.}\ \bibnamefont {Regal}},\
  }\href@noop {} {\bibfield  {journal} {\bibinfo  {journal} {Science},\
  }\textbf {\bibinfo {volume} {339}},\ \bibinfo {pages} {801} (\bibinfo {year}
  {2013})}\BibitemShut {NoStop}%
\bibitem [{\citenamefont {Marquardt}\ \emph {et~al.}(2007)\citenamefont
  {Marquardt}, \citenamefont {Chen}, \citenamefont {Clerk},\ and\ \citenamefont
  {Girvin}}]{Marquardt2007}%
  \BibitemOpen
  \bibfield  {author} {\bibinfo {author} {\bibfnamefont {F.}~\bibnamefont
  {Marquardt}}, \bibinfo {author} {\bibfnamefont {J.~P.}\ \bibnamefont {Chen}},
  \bibinfo {author} {\bibfnamefont {A.~A.}\ \bibnamefont {Clerk}}, \ and\
  \bibinfo {author} {\bibfnamefont {S.~M.}\ \bibnamefont {Girvin}},\
  }\href@noop {} {\bibfield  {journal} {\bibinfo  {journal} {Phys. Rev.
  Lett.},\ }\textbf {\bibinfo {volume} {99}},\ \bibinfo {pages} {093902}
  (\bibinfo {year} {2007})}\BibitemShut {NoStop}%
\bibitem [{\citenamefont {Wilson-Rae}\ \emph {et~al.}(2007)\citenamefont
  {Wilson-Rae}, \citenamefont {Nooshi}, \citenamefont {Zwerger},\ and\
  \citenamefont {Kippenberg}}]{Wilson-Rae2007}%
  \BibitemOpen
  \bibfield  {author} {\bibinfo {author} {\bibfnamefont {I.}~\bibnamefont
  {Wilson-Rae}}, \bibinfo {author} {\bibfnamefont {N.}~\bibnamefont {Nooshi}},
  \bibinfo {author} {\bibfnamefont {W.}~\bibnamefont {Zwerger}}, \ and\
  \bibinfo {author} {\bibfnamefont {T.}~\bibnamefont {Kippenberg}},\
  }\href@noop {} {\bibfield  {journal} {\bibinfo  {journal} {Phys. Rev.
  Lett.},\ }\textbf {\bibinfo {volume} {99}},\ \bibinfo {pages} {093901}
  (\bibinfo {year} {2007})}\BibitemShut {NoStop}%
\bibitem [{\citenamefont {Genes}\ \emph {et~al.}(2008)\citenamefont {Genes},
  \citenamefont {Vitali}, \citenamefont {Tombesi}, \citenamefont {Gigan},\ and\
  \citenamefont {Aspelmeyer}}]{Genes2008}%
  \BibitemOpen
  \bibfield  {author} {\bibinfo {author} {\bibfnamefont {C.}~\bibnamefont
  {Genes}}, \bibinfo {author} {\bibfnamefont {D.}~\bibnamefont {Vitali}},
  \bibinfo {author} {\bibfnamefont {P.}~\bibnamefont {Tombesi}}, \bibinfo
  {author} {\bibfnamefont {S.}~\bibnamefont {Gigan}}, \ and\ \bibinfo {author}
  {\bibfnamefont {M.}~\bibnamefont {Aspelmeyer}},\ }\href@noop {} {\bibfield
  {journal} {\bibinfo  {journal} {Phys. Rev. A},\ }\textbf {\bibinfo {volume}
  {77}},\ \bibinfo {pages} {033804} (\bibinfo {year} {2008})}\BibitemShut
  {NoStop}%
\bibitem [{\citenamefont {Chan}\ \emph {et~al.}(2011)\citenamefont {Chan},
  \citenamefont {Alegre}, \citenamefont {Safavi-Naeini}, \citenamefont {Hill},
  \citenamefont {Krause}, \citenamefont {Gr\"{o}blacher}, \citenamefont
  {Aspelmeyer},\ and\ \citenamefont {Painter}}]{Chan2011}%
  \BibitemOpen
  \bibfield  {author} {\bibinfo {author} {\bibfnamefont {J.}~\bibnamefont
  {Chan}}, \bibinfo {author} {\bibfnamefont {T.~P.~M.}\ \bibnamefont {Alegre}},
  \bibinfo {author} {\bibfnamefont {A.~H.}\ \bibnamefont {Safavi-Naeini}},
  \bibinfo {author} {\bibfnamefont {J.~T.}\ \bibnamefont {Hill}}, \bibinfo
  {author} {\bibfnamefont {A.}~\bibnamefont {Krause}}, \bibinfo {author}
  {\bibfnamefont {S.}~\bibnamefont {Gr\"{o}blacher}}, \bibinfo {author}
  {\bibfnamefont {M.}~\bibnamefont {Aspelmeyer}}, \ and\ \bibinfo {author}
  {\bibfnamefont {O.}~\bibnamefont {Painter}},\ }\href@noop {} {\bibfield
  {journal} {\bibinfo  {journal} {Nature (London)},\ }\textbf {\bibinfo
  {volume} {478}},\ \bibinfo {pages} {89} (\bibinfo {year} {2011})}\BibitemShut
  {NoStop}%
\bibitem [{\citenamefont {Verhagen}\ \emph {et~al.}(2012)\citenamefont
  {Verhagen}, \citenamefont {Del\'eglise}, \citenamefont {Weis}, \citenamefont
  {Schliesser},\ and\ \citenamefont {Kippenberg}}]{Verhagen2012}%
  \BibitemOpen
  \bibfield  {author} {\bibinfo {author} {\bibfnamefont {E.}~\bibnamefont
  {Verhagen}}, \bibinfo {author} {\bibfnamefont {S.}~\bibnamefont
  {Del\'eglise}}, \bibinfo {author} {\bibfnamefont {S.}~\bibnamefont {Weis}},
  \bibinfo {author} {\bibfnamefont {A.}~\bibnamefont {Schliesser}}, \ and\
  \bibinfo {author} {\bibfnamefont {T.~J.}\ \bibnamefont {Kippenberg}},\
  }\href@noop {} {\bibfield  {journal} {\bibinfo  {journal} {Nature (London)},\
  }\textbf {\bibinfo {volume} {482}},\ \bibinfo {pages} {63} (\bibinfo {year}
  {2012})}\BibitemShut {NoStop}%
\bibitem [{\citenamefont {Uhlmann}(1976)}]{Uhlmann1976}%
  \BibitemOpen
  \bibfield  {author} {\bibinfo {author} {\bibfnamefont {A.}~\bibnamefont
  {Uhlmann}},\ }\href@noop {} {\bibfield  {journal} {\bibinfo  {journal} {Rep.
  Math. Phys},\ }\textbf {\bibinfo {volume} {9}},\ \bibinfo {pages} {273}
  (\bibinfo {year} {1976})}\BibitemShut {NoStop}%
\bibitem [{\citenamefont {Jozsa}(1994)}]{Jozsa1994}%
  \BibitemOpen
  \bibfield  {author} {\bibinfo {author} {\bibfnamefont {R.}~\bibnamefont
  {Jozsa}},\ }\href@noop {} {\bibfield  {journal} {\bibinfo  {journal} {J. Mod.
  Opt.},\ }\textbf {\bibinfo {volume} {41}},\ \bibinfo {pages} {2315} (\bibinfo
  {year} {1994})}\BibitemShut {NoStop}%
\bibitem [{\citenamefont {Wang}\ \emph {et~al.}(2008)\citenamefont {Wang},
  \citenamefont {Yu},\ and\ \citenamefont {Yi}}]{Wang2008}%
  \BibitemOpen
  \bibfield  {author} {\bibinfo {author} {\bibfnamefont {X.}~\bibnamefont
  {Wang}}, \bibinfo {author} {\bibfnamefont {C.-S.}\ \bibnamefont {Yu}}, \ and\
  \bibinfo {author} {\bibfnamefont {X.}~\bibnamefont {Yi}},\ }\href@noop {}
  {\bibfield  {journal} {\bibinfo  {journal} {Phys. Lett. A},\ }\textbf
  {\bibinfo {volume} {373}},\ \bibinfo {pages} {58} (\bibinfo {year}
  {2008})}\BibitemShut {NoStop}%
\bibitem [{\citenamefont {Nunnenkamp}\ \emph {et~al.}(2010)\citenamefont
  {Nunnenkamp}, \citenamefont {B{\o}rkje}, \citenamefont {Harris},\ and\
  \citenamefont {Girvin}}]{Nunnenkamp2010}%
  \BibitemOpen
  \bibfield  {author} {\bibinfo {author} {\bibfnamefont {A.}~\bibnamefont
  {Nunnenkamp}}, \bibinfo {author} {\bibfnamefont {K.}~\bibnamefont
  {B{\o}rkje}}, \bibinfo {author} {\bibfnamefont {J.~G.~E.}\ \bibnamefont
  {Harris}}, \ and\ \bibinfo {author} {\bibfnamefont {S.~M.}\ \bibnamefont
  {Girvin}},\ }\href@noop {} {\bibfield  {journal} {\bibinfo  {journal} {Phys.
  Rev. A},\ }\textbf {\bibinfo {volume} {82}},\ \bibinfo {pages} {021806(R)}
  (\bibinfo {year} {2010})}\BibitemShut {NoStop}%
\bibitem [{\citenamefont {Kennedy}\ and\ \citenamefont
  {Walls}(1988)}]{Kennedy1988}%
  \BibitemOpen
  \bibfield  {author} {\bibinfo {author} {\bibfnamefont {T.}~\bibnamefont
  {Kennedy}}\ and\ \bibinfo {author} {\bibfnamefont {D.}~\bibnamefont
  {Walls}},\ }\href@noop {} {\bibfield  {journal} {\bibinfo  {journal} {Phys.
  Rev. A},\ }\textbf {\bibinfo {volume} {37}},\ \bibinfo {pages} {152}
  (\bibinfo {year} {1988})}\BibitemShut {NoStop}%
\bibitem [{\citenamefont {Kim}\ and\ \citenamefont
  {Buz\v{e}k}(1992)}]{Kim1992}%
  \BibitemOpen
  \bibfield  {author} {\bibinfo {author} {\bibfnamefont {M.}~\bibnamefont
  {Kim}}\ and\ \bibinfo {author} {\bibfnamefont {V.}~\bibnamefont
  {Buz\v{e}k}},\ }\href@noop {} {\bibfield  {journal} {\bibinfo  {journal}
  {Phys. Rev. A},\ }\textbf {\bibinfo {volume} {46}},\ \bibinfo {pages} {4239}
  (\bibinfo {year} {1992})}\BibitemShut {NoStop}%
\bibitem [{\citenamefont {Brune}\ \emph {et~al.}(1996)\citenamefont {Brune},
  \citenamefont {Hagley}, \citenamefont {Dreyer}, \citenamefont {Ma\^itre},
  \citenamefont {Maali}, \citenamefont {Wunderlich}, \citenamefont {Raimond},\
  and\ \citenamefont {Haroche}}]{Brune1996}%
  \BibitemOpen
  \bibfield  {author} {\bibinfo {author} {\bibfnamefont {M.}~\bibnamefont
  {Brune}}, \bibinfo {author} {\bibfnamefont {E.}~\bibnamefont {Hagley}},
  \bibinfo {author} {\bibfnamefont {J.}~\bibnamefont {Dreyer}}, \bibinfo
  {author} {\bibfnamefont {X.}~\bibnamefont {Ma\^itre}}, \bibinfo {author}
  {\bibfnamefont {A.}~\bibnamefont {Maali}}, \bibinfo {author} {\bibfnamefont
  {C.}~\bibnamefont {Wunderlich}}, \bibinfo {author} {\bibfnamefont {J.~M.}\
  \bibnamefont {Raimond}}, \ and\ \bibinfo {author} {\bibfnamefont
  {S.}~\bibnamefont {Haroche}},\ }\href@noop {} {\bibfield  {journal} {\bibinfo
   {journal} {Phys. Rev. Lett.},\ }\textbf {\bibinfo {volume} {77}},\ \bibinfo
  {pages} {4887} (\bibinfo {year} {1996})}\BibitemShut {NoStop}%
\bibitem [{\citenamefont {Genoni}\ \emph {et~al.}(2013)\citenamefont {Genoni},
  \citenamefont {Palma}, \citenamefont {Tufarelli}, \citenamefont {Olivares},
  \citenamefont {Kim},\ and\ \citenamefont {Paris}}]{Genoni2013}%
  \BibitemOpen
  \bibfield  {author} {\bibinfo {author} {\bibfnamefont {M.~G.}\ \bibnamefont
  {Genoni}}, \bibinfo {author} {\bibfnamefont {M.}~\bibnamefont {Palma}},
  \bibinfo {author} {\bibfnamefont {T.}~\bibnamefont {Tufarelli}}, \bibinfo
  {author} {\bibfnamefont {S.}~\bibnamefont {Olivares}}, \bibinfo {author}
  {\bibfnamefont {M.}~\bibnamefont {Kim}}, \ and\ \bibinfo {author}
  {\bibfnamefont {M.}~\bibnamefont {Paris}},\ }\href@noop {} {\bibfield
  {journal} {\bibinfo  {journal} {Phys. Rev. A},\ }\textbf {\bibinfo {volume}
  {87}},\ \bibinfo {pages} {062104} (\bibinfo {year} {2013})}\BibitemShut
  {NoStop}%
\bibitem [{\citenamefont {Palma}\ \emph {et~al.}(2013)\citenamefont {Palma},
  \citenamefont {Stammers}, \citenamefont {Genoni}, \citenamefont {Tufarelli},
  \citenamefont {Olivares}, \citenamefont {Kim},\ and\ \citenamefont
  {Paris}}]{Palma2013}%
  \BibitemOpen
  \bibfield  {author} {\bibinfo {author} {\bibfnamefont {M.~L.}\ \bibnamefont
  {Palma}}, \bibinfo {author} {\bibfnamefont {J.}~\bibnamefont {Stammers}},
  \bibinfo {author} {\bibfnamefont {M.~G.}\ \bibnamefont {Genoni}}, \bibinfo
  {author} {\bibfnamefont {T.}~\bibnamefont {Tufarelli}}, \bibinfo {author}
  {\bibfnamefont {S.}~\bibnamefont {Olivares}}, \bibinfo {author}
  {\bibfnamefont {M.~S.}\ \bibnamefont {Kim}}, \ and\ \bibinfo {author}
  {\bibfnamefont {M.~G.~A.}\ \bibnamefont {Paris}},\ }\href@noop {} {\bibfield
  {journal} {\bibinfo  {journal} {arXiv:1309.4221v1}} (\bibinfo {year}
  {2013})}\BibitemShut {NoStop}%
\bibitem [{\citenamefont {D'Ariano}\ \emph {et~al.}(1994)\citenamefont
  {D'Ariano}, \citenamefont {Macchiavello},\ and\ \citenamefont
  {Paris}}]{Dariano1994}%
  \BibitemOpen
  \bibfield  {author} {\bibinfo {author} {\bibfnamefont {G.}~\bibnamefont
  {D'Ariano}}, \bibinfo {author} {\bibfnamefont {C.}~\bibnamefont
  {Macchiavello}}, \ and\ \bibinfo {author} {\bibfnamefont {M.}~\bibnamefont
  {Paris}},\ }\href@noop {} {\bibfield  {journal} {\bibinfo  {journal} {Phys.
  Rev. A},\ }\textbf {\bibinfo {volume} {50}},\ \bibinfo {pages} {4298–4302}
  (\bibinfo {year} {1994})}\BibitemShut {NoStop}%
\bibitem [{\citenamefont {Vitali}\ \emph {et~al.}(2007)\citenamefont {Vitali},
  \citenamefont {Gigan}, \citenamefont {Ferreira}, \citenamefont {Bohm},
  \citenamefont {Tombesi}, \citenamefont {Guerreiro}, \citenamefont {Vedral},
  \citenamefont {Zeilinger},\ and\ \citenamefont {Aspelmeyer}}]{Vitali2007}%
  \BibitemOpen
  \bibfield  {author} {\bibinfo {author} {\bibfnamefont {D.}~\bibnamefont
  {Vitali}}, \bibinfo {author} {\bibfnamefont {S.}~\bibnamefont {Gigan}},
  \bibinfo {author} {\bibfnamefont {A.}~\bibnamefont {Ferreira}}, \bibinfo
  {author} {\bibfnamefont {H.~R.}\ \bibnamefont {Bohm}}, \bibinfo {author}
  {\bibfnamefont {P.}~\bibnamefont {Tombesi}}, \bibinfo {author} {\bibfnamefont
  {A.}~\bibnamefont {Guerreiro}}, \bibinfo {author} {\bibfnamefont
  {V.}~\bibnamefont {Vedral}}, \bibinfo {author} {\bibfnamefont
  {A.}~\bibnamefont {Zeilinger}}, \ and\ \bibinfo {author} {\bibfnamefont
  {M.}~\bibnamefont {Aspelmeyer}},\ }\href@noop {} {\bibfield  {journal}
  {\bibinfo  {journal} {Phys. Rev. Lett.},\ }\textbf {\bibinfo {volume} {98}},\
  \bibinfo {pages} {030405} (\bibinfo {year} {2007})}\BibitemShut {NoStop}%
\bibitem [{\citenamefont {Vanner}\ \emph {et~al.}(2011)\citenamefont {Vanner},
  \citenamefont {Pikovski}, \citenamefont {Cole}, \citenamefont {Kim},
  \citenamefont {\v{C} Brukner}, \citenamefont {Hammerer}, \citenamefont
  {Milburn},\ and\ \citenamefont {Aspelmeyer}}]{Vanner2011}%
  \BibitemOpen
  \bibfield  {author} {\bibinfo {author} {\bibfnamefont {M.~R.}\ \bibnamefont
  {Vanner}}, \bibinfo {author} {\bibfnamefont {I.}~\bibnamefont {Pikovski}},
  \bibinfo {author} {\bibfnamefont {G.~D.}\ \bibnamefont {Cole}}, \bibinfo
  {author} {\bibfnamefont {M.~S.}\ \bibnamefont {Kim}}, \bibinfo {author}
  {\bibnamefont {\v{C} Brukner}}, \bibinfo {author} {\bibfnamefont
  {K.}~\bibnamefont {Hammerer}}, \bibinfo {author} {\bibfnamefont {G.~J.}\
  \bibnamefont {Milburn}}, \ and\ \bibinfo {author} {\bibfnamefont
  {M.}~\bibnamefont {Aspelmeyer}},\ }\href@noop {} {\bibfield  {journal}
  {\bibinfo  {journal} {Proc. Nat. Acad. Sci},\ }\textbf {\bibinfo {volume}
  {108}},\ \bibinfo {pages} {16182} (\bibinfo {year} {2011})}\BibitemShut
  {NoStop}%
\end{thebibliography}%

\end{document}